\documentclass[usenatbib]{mn2e}

\usepackage[english]{babel}
\usepackage[latin1]{inputenc}
\usepackage[T1]{fontenc}

\usepackage{amsmath}
\usepackage{bbold}
\usepackage{graphicx}
\usepackage{amsfonts}
\usepackage{amssymb}
\usepackage{hyperref}
\usepackage{latexsym}

\newcommand{\mnras}{{MNRAS }}

\newcommand{\aj}{{AJ }}
\newcommand{\apj}{{ApJ }}
\newcommand{\apjs}{{ApJS }}

\newcommand{\be}{\begin{equation}}
\newcommand{\ee}{\end{equation}}
\newcommand{\bea}{\begin{eqnarray}}
\newcommand{\eea}{\end{eqnarray}}
\newcommand{\no}{\nonumber}
\newcommand{\dk}{\frac{d^3 k}{(2\pi)^3}}
\newcommand{\dkp}{\frac{d^3 k'}{(2\pi)^3}}
\newcommand{\brak}{\langle \vec{k} |}
\newcommand{\ketk}{| \vec{k} \rangle}
\newcommand{\brax}{\langle \vec{x} |}
\newcommand{\ketx}{| \vec{x} \rangle}

\newcommand{\ketkp}{| \vec{k}' \rangle}

\begin{document}

\title[The full Fisher matrix for galaxy surveys]{The full Fisher matrix for galaxy surveys}

\author[L.~Raul~Abramo]{L.~Raul Abramo $^{1,2,3}$
\\
$^1$ Department of Physics \& Astronomy, University of Pennsylvania, 
Philadelphia, PA, 19104 \\
$^2$ Department of Astrophysical Sciences, Princeton University, 
Peyton Hall, Princeton, NJ, 08544 \\
$^3$ Departamento de Física Matemática,
Instituto de F\'{\i}sica, Universidade de S\~ao Paulo, 
CP 66318, CEP 05314-970 São Paulo, Brazil}

\maketitle

\begin{abstract}
Starting from the Fisher matrix for counts in cells,
I derive the full Fisher matrix for surveys of multiple 
tracers of large-scale structure. The key 
step is the ``classical approximation'', which allows to write
the inverse of the covariance of the galaxy counts 
in terms of
the naive matrix inverse of the covariance in a mixed 
position-space and Fourier-space basis.
I then compute the Fisher matrix for the power spectrum in bins of the
three-dimensional wavenumber $\vec{k}$; the Fisher matrix for functions 
of position $\vec{x}$ (or redshift $z$) such as the linear bias of the tracers 
and/or the growth function; and the cross-terms of the Fisher matrix that 
expresses the correlations between estimations of the power spectrum
and estimations of the bias. 
When the bias and growth function are fully specified, and the Fourier-space 
bins are large enough 
that the covariance between them can be neglected, the Fisher matrix for
the power spectrum reduces to the widely used result that was first derived 
by Feldman, Kaiser and Peacock (1994). 
Assuming isotropy, 
a fully analytical calculation of the 
Fisher matrix in the classical approximation can be 
performed in the case of a constant-density, volume-limited survey.
\end{abstract}



\section{Introduction}

With the growing relevance, cost and complexity of galaxy surveys
[\cite{York:2000gk,cole_2df_2005,Abbott:2005bi,scoville_cosmic_2007,adelman-mccarthy_sdss_2008,adelman-mccarthy_sixth_2008,PAN-STARRS,Benitez:2008fs,BOSS,LSST:2009pq,SUMIRE,2011MNRAS.415.2876B}], 
the scientific potential of these probes must be accurately projected. Since 
that potential is usually expressed in terms of constraints on the currently favored 
theoretical models and their parameters, forecasting those constraints 
is a critical part of the design and justification of any new survey  [\cite{Albrecht09}].

One of the most interesting functions that one wishes to constrain with these galaxy surveys
is the power spectrum $P(k)$, as well as its sub-products such as the
baryon acoustic oscillations [\cite{eisenstein_cosmic_1998,blake_probing_2003,
seo_probing_2003}].
If our theories about the origin of structure in the Universe are correct 
[\cite{Mukhanov:05,Peter:2009zzc}], the power 
spectrum should be given by the expectation
value $\langle \delta (\vec{k}) \delta^*(\vec{k}') \rangle = (2\pi)^3 P(k) 
\delta_D (\vec{k}-\vec{k}')$, where $\delta(\vec{k})$ is (in linear 
perturbation theory) a nearly
scale-invariant Gaussian random field corresponding to the Fourier transform 
of the density fluctuation contrast, $\delta(\vec{x}) =
\delta \rho(\vec{x})/\rho(\vec{x})$, and
$\delta_D$ is the Dirac delta function (in this paper position space is always expressed
in terms of comoving coordinates $\vec{x}$).

However, in the absence of information about gravitational lensing, which
can probe directly the total masses of halos, all that we are able 
to measure with galaxy surveys are the positions of individual 
galaxies and other (biased) tracers of the underlying large-scale structure, 
over a finite volume and with a limited spatial accuracy. 
Hence, all estimates for the power 
spectrum are based on the two-point correlation functions of these tracers, 
which are themselves proportional to the Fourier transform of the power spectrum
-- the constants of proportionality being the bias of each type of tracer.

There are many theoretical and practical problems involved in the estimation 
of a Fourier space function, $P(k)$, from (imperfect) measurements of its 
counterpart in position space, the two-point correlation function 
$\xi(x)$ [see \cite{1994ApJ...424..569B} for the issues that arise when 
trying to estimate $\xi$ directly from the data)].
First, the mechanism whereby one or more galaxies
appear at the peak of a local density field leads to shot noise -- i.e., 
statistical fluctuations typical of point processes, usually assumed 
to be of a Poisson nature.
Second, the finite volume mapped in a real survey leads to sample
(or cosmic) variance, which limits the accuracy with which 
we can estimate the power spectrum for any given mode $k$
-- or, equivalently, $\xi$ at a scale $x$.
These concerns imply that, when building an estimator for the power spectrum,
one must weigh each galaxy pair in such a way to minimize the variance of
that estimator, while ensuring that it remains unbiased.
Although the present work aims to establish a more firm basis for the basic tools 
used for estimation and forecast of parameters from galaxy surveys, it is
important to recognize that in practice the ``real world'' problems can be
even harder to tackle -- for reviews, see \cite{HamiltonRev05a,HamiltonRev05b}.

In a seminal paper Feldman, Kaiser and Peacock [\cite{1994ApJ...426...23F}] 
(hereafter, FKP) showed that there is an ``optimal'' estimator of the power spectrum,
in the sense that the combined contributions from shot noise and sample variance 
to the covariance of that estimator are minimized.
That estimator takes the form of a weighting function for pairs of galaxies 
which depends on the Fourier mode of the spectrum that is being estimated,
$U(\vec{x},\vec{k}) = \bar{n}(\vec{x}) b^2 (\vec{x}) D^2 (x)  P(\vec{k}) /
[1 +  \bar{n}(\vec{x}) b^2 (\vec{x}) D^2 (x) P(\vec{k}) ] $, where 
$\bar{n}$ is the average volumetric density of galaxies
(sometimes also referred to in the literature, somewhat confusingly, 
as the selection function); $b$ is the bias of
those galaxies (which we will assume to be linear and deterministic); 
and $D(x)$ is the linear matter growth function at the redshift $z(x)$. 
By factoring out the linear growth function (which is normalized to $D=1$ at
$z=0$) , I am implicitly taking $P(\vec{k})$ to mean the linear matter power 
spectrum, normalized at $z=0$.
Both the growth function and the power spectrum depend on a number
of fundamental cosmological parameters ($h$, $\Omega_m$, $\Omega_\Lambda$,
$\sigma_8$, $w$, $\gamma$, etc.), which is what we ultimately would like
to constrain.

In a nutshell, the FKP result implies that the contribution from galaxies inside the 
volume $dV_x$ to the variance of the power spectrum estimated over some 
Fourier-space volume $dV_k=4\pi k^2 dk/(2\pi)^3$ (i.e., the {\it bandpower} at $dV_k$), 
is given approximately by $ \sigma_{P}^2 = (\Delta P/P)^2 = 2 /(U^{2} \, dV_k \, dV_x)$. 
A related quantity of interest is the effective volume [\cite{Tegmark_Surveys_1997}]
for the mode $k$, defined as $V_{\rm eff}(k) =  \int dV_x \, U^{2}$.
These results were later generalized to include the case where different
tracers of large-scale structure (i.e., tracers with different biases) are used jointly 
to constrain the power spectrum [\cite{Percival:2003pi,White:2008jy,McDonald:2008sh}].

The FKP formula has an intuitive interpretation in terms of functions of phase space.
The density of information contained in a phase space cell centered on $(\vec{k},\vec{x})$ 
is determined by two densities: the effective biased density of galaxies, 
$N \equiv \bar{n} \, b^2 D^2$, 
which (under the assumption of linear and deterministic bias) is a function
of position $\vec{x}$, and the linear spectrum $P(\vec{k})$, which is the density of modes 
in Fourier space. The adimensional function $\frac12 U^2=\frac12 [N P/ (1 + N P)]^2$ 
can then be interpreted as some kind of density of information in phase space.

In a series of elegant papers, A. Hamilton, M. Tegmark and collaborators 
[\cite{Hamilton_97_I,Hamilton_97_II,Tegmark_Surveys_1997,hamilton_linear_1997,1998ApJ...499..555T}] showed
that the FKP formulas can be derived from the Fisher matrix of galaxy counts
in cells (i.e., when the spatial cells are the pixels), under the assumption of gaussianity. 
In that case the Fisher matrix for the power spectrum can be written in terms of the 
pixel-space covariance $C(\delta_x,\delta_{x'}) = \langle  \delta_x \delta_x^{t} \rangle$ 
and its derivatives with respect to the bandpowers $P_i=P(k_i)$ of the 
(fiducial) power spectrum in the usual [\cite{1996ApJ...465...34V}] way:
$F_{ij} = 1/2 \, {\rm Tr} \, [ \, C^{-1} \, C_{,i} \, C^{-1} \, C_{,j} \, ]$. It was then shown 
[\cite{Hamilton_97_I,1998ApJ...499..555T}] that, under some assumptions and after
some approximations, the Fisher matrix reduces to the FKP formula for the 
(inverse) variance of the power spectrum. 
This important result provides the connection between forecasts in a best-case scenario
(which, because of the Cram\'er-Rao bound, are given by the Fisher matrix),
and the estimation of the power spectrum from real data -- e.g., the Fisher matrix-based 
quadratic methods [\cite{1998ApJ...499..555T}]
and/or dimensional reduction
methods employing pseudo-Karhune-Loeve (pKL) eigenmodes 
(which become, in effect, the pixels).
These methods have been extensively employed in the analysis of the power 
spectrum in the SDSS, and are reviewed by \cite{Tegmark:2003uf,Tegmark:2006az}.

It is well known, however, that the FKP effective volume suffers from some limitations.

First, the FKP Fisher matrix associated with $V_{eff}(k)$ is purely diagonal,
$F_{FKP} (\vec{k},\vec{k}') \sim V_{eff}(\vec{k}) \, \delta_{\vec{k}, \vec{k}'}$,
which means that it neglects the covariance between the estimates of the  
bandpowers $P(\vec{k})$ and $P(\vec{k}')$. This not only overstates
the constraining power of the Fisher matrix, but it also does not allow us to
estimate the optimal size of the bins or, equivalently, to compute
the principal components of the full matrix. Knowledge of the
full Fisher matrix would be useful to improve the forecasts of constraints
on cosmological parameters, to obtain the minimal size of the $k$ bins,
and even to inform the choice of pKL modes.

Second, the effective volume does not take proper account of long-range 
correlations. In order to better appreciate this deficiency,
consider the pathological case of a galaxy catalog that is formed by two 
disjointed volumes, $V_1$ and $V_2$.
For simplicity, assume that the average effective number density of galaxies is the same 
($N_0$) in both volumes. The FKP 
formula then tells us that the variance of the power spectrum at the scale $k$ that 
can be estimated with that catalog is:
\begin{eqnarray}
\label{Eq:Disj}
\sigma_{P}^{-2}(k) &=&  \frac12 \,dV_k \, \int U^{2}  dV_x  
\\ \nonumber
&=& \frac12  \, dV_k \, \left[ \frac{N_0 P(k)}{1+N_0 P(k)} \right]^2 \times (V_1 +V_2) \; . 
\end{eqnarray}
We recognize this as the sum of the diagonal terms of the Fisher matrices for 
the galaxies in $V_1$ and that for the galaxies in $V_2$. But there is no cross-correlation 
term, meaning that the information residing in the correlation 
between any galaxy in $V_1$ and any other in $V_2$ has been somehow neglected in that approximation \footnote{I would like to thank Ravi Sheth for bringing this puzzle to my attention.}.

These problems arise out of what Hamilton has called the
``classical approximation'' 
[\cite{Hamilton_97_I,hamilton_linear_1997}], whereby only
galaxies in the same shell in position space, and only power spectrum estimates 
in the same shell in Fourier space, are allowed to have non-vanishing correlations. 
The term ``classical limit'' is inherited from the language of quantum 
mechanics, and that toolbox turns out to be useful in the context of the statistics
of galaxy surveys \footnote{This paper is heavily indebted to 
ideas and notation set forth in A. Hamilton's papers [\cite{1996MNRAS.278...73H,Hamilton_97_I,Hamilton_97_II,hamilton_linear_1997}],
who was (to my knowledge) the first to make extensive use of the analogy between
stochasticity in galaxy surveys and the language of quantum mechanics, e.g., 
treating the two-point correlation function and the power spectrum as a single 
``operator'' expressed in two different basis (position- and Fourier-space, 
respectively).}. The language and concepts of quantum
mechanics are convenient in this context because some objects of interest can 
be diagonalized in one basis, but not the other: e.g., the linear power spectrum is diagonal 
(at least in standard linear theory) only in the Fourier basis, while the shot noise term in the 
covariance of galaxy counts is diagonal only in the position-space basis -- hence,
in that sense, these two operators do not commute. 
The covariance of galaxy counts, however, is not
diagonal in either one of these basis, which is the main complicating factor.

The FKP result follows from making two distinct approximations in the
Fisher matrix of galaxy surveys:
i) the first step (the ``classical approximation'') 
is to take all operators (such as the correlation function or shot noise) to be classical, 
and therefore commuting with each other;
ii) the second step consists in taking the limit whereby the phase space 
window functions 
$\exp { \left[ i (\vec{k}-\vec{k}') \cdot (\vec{x} -\vec{x}') \right]}
\rightarrow (2\pi)^3 \delta_D (\vec{k}-\vec{k}') \delta_D (\vec{x}-\vec{x}') $. 
This limit is basically a stationary phase (SP) approximation.
%

\vskip 0.3cm

This paper shows how to obtain an exact, but formal, expression for the full Fisher matrix
of galaxy surveys. It also shows hot compute the full Fisher matrix
in the classical approximation -- but without having to make use of the SP approximation.
The calculation of the Fisher matrix in the classical limit relies on the use of a 
phase-space basis (in both position space and Fourier space), which allows the 
inversion of the pixel-space covariance in that approximation.
I compute not only the Fisher matrix for the estimation of the power
spectrum on bins of the Fourier modes $k$, but also the Fisher matrix
for the estimation of functions of redshift (such as the bias of each tracer and/or the growth
function), as well as the cross terms of the Fisher matrix which express the 
correlation between estimations of the power spectrum as a function of $k$ and 
estimations of bias (and/or growth function) as a function of $z$.
Readers uninterested in the details of the calculation can skip to
the end of Section \ref{S:QM}, where the main results of this paper are 
summarized by Eqs. (\ref{Eq:U})-(\ref{Eq:FMixPN}), as well as their
classical limits, Eqs. (\ref{Eq:GFPclass})-(\ref{Eq:GFNclass}).

\vskip 0.3cm

This paper is organized as follows. In Section \ref{S:F} the Fisher matrix for an 
arbitrary survey of multiple types of tracers of large-scale structure is 
derived from the covariance of galaxy counts using standard notation. 
In Section \ref{S:QM} I examine the Fisher matrix from the perspective of 
objects borrowed from quantum mechanics -- operators, basis vectors, states, etc.
Starting from the covariance matrix in phase space and its na\"ive inverse, 
I derive the full Fisher matrix for galaxy surveys, and show that it has all the right properties 
-- including the fact that, of course, it reduces to the FKP formula after taking both the 
classical and the SP approximations. 
Still in Section \ref{S:QM}, I compute the Fisher matrix for the bias 
and/or growth function, as well as the terms of the full Fisher matrix which mix the 
estimation of the power spectrum with the estimation of position-space functions
from the same galaxy survey.
Finally, in Section \ref{S:A} I consider, as an application, an isotropic survey with
effective number density $N(x)$ and an isotropic power spectrum $P(k)$. 
When $N(x)$ is given by a top-hat profile, there exists an analytical solution for
the Fisher matrix in the classical limit (but without having to assume the SP
approximation). That analytical solution shows that cross-correlations
between different bandpowers can arise if the Fourier-space bins are too 
small -- a problem that may affect some recent analyses of large-scale
structure [\cite{2010MNRAS.401.2148P}].
In that Section I also derive an analytical expression for the Fisher matrix
which measures the information contained in the cross-correlation between two 
species of tracers with top-hat density profiles.
In particular, I show that in this example the Fisher matrix can be expressed in terms of 
phase space window functions which are basically identical to the Fisher matrix 
that follows from expressions found in \cite{Hamilton_97_I,Hamilton_97_II}.
I present, in an Appendix, a semi-analytical formula for the Fisher 
matrix in the case of a galaxy survey with an arbitrary distribution of any 
number of different tracers of large-scale structure.

\vskip 0.3cm 

For this purposes of this paper I will only work in position (real) space, but the 
generalization to redshift space is straightforward: the power spectrum, 
in particular, inherits the redshift distortions and the associated 
dependence on the direction of the modes,
$P(k) \to P^s(k_\perp,k_{||})$. I also do not fully explore the Fisher matrix for
position-dependent degrees of freedom such as the bias and growth
function -- that will be the subject of a forthcoming paper.



\section{The covariance matrix of galaxy counts and the classical approximation}
\label{S:F}

The basic object in the construction of the Fisher matrix is the covariance matrix
for galaxy counts. I will consider many different species of tracers (e.g., 
red galaxies [\cite{tegmark_3d_2003,Tegmark:1996bz}], 
blue galaxies [\cite{2002MNRAS.332..827N,Tegmark:2003ud}], 
emission-line galaxies [\cite{2011MNRAS.415.2876B}], 
neutral H regions probed by quasar absorption lines [\cite{Seljak:2004xh}], 
quasars [\cite{2011arXiv1108.1198S,2011arXiv1108.2657A}], etc.), with 
mean number densities and linear biases given by 
$\bar{n}_\mu (\vec{x})$ and $b_\mu (\vec{x})$, respectively, 
where greek indices $\mu = 1 \ldots N_t$ denote the different types of tracers.
The two-point correlation function between the counts of any two types of tracers is
given by:
\bea
\label{Eq:2PCF}
\xi_{\mu \nu} 
&=& b_\mu (\vec{x}) D(\vec{x}) \, b_\nu (\vec{y}) D(\vec{y}) \,
\xi (\vec{x} - \vec{y}) 
\\ \nonumber
& \equiv & B_\mu (\vec{x}) \, B_\nu (\vec{y}) 
\int \dk e^{-i \vec{k} \cdot (\vec{x} - \vec{y})} P(\vec{k}) \; ,
\eea
where we have included the matter growth function $D(\vec{x}) = D(z[x])$ into the
definition of an effective bias $B_\mu \equiv b_\mu D$ (which is assumed linear
and deterministic), and $P(\vec{k})$ is the
{\em linear} matter power spectrum. Because the power spectrum is the Fourier 
transform of the real function $\xi$, it obeys $P^*(\vec{k}) = P(-\vec{k})$. If
isotropy holds, the spectrum is a (real) function of $k=|\vec{k}|$, but I will not assume
this until we come to Section IV. Notice that $\xi_{\mu \nu} (\vec{x},\vec{y})
= \xi_{\nu \mu} (\vec{y},\vec{x})$, but $ \xi_{\mu \nu} (\vec{x},\vec{y}) \neq \xi_{\mu \nu} (\vec{y},\vec{x})$. 

Finding and mapping individual objects is basically a point process, 
subject therefore to stochasticity (shot noise, in this context).
Cosmologists usually make the simplest possible assumption and take a Poisson 
distribution for the shot noise of counts in cells -- although this may be an overestimate, 
particularly if one counts halos instead of galaxies [\cite{2011MNRAS.412..995C}]. 
In that case, the covariance of galaxy counts can be expressed as:
\bea
\label{Eq:COV}
C_{\mu \nu} (\vec{x},\vec{x}') &=& \xi_{\mu \nu} (\vec{x},\vec{x}') + \frac{1}{\bar{n}_\mu (\vec{x})} \,
\delta_{\mu \nu} \, \delta_D (\vec{x} - \vec{x}') 
\\ \nonumber
&=& \int \dk \, e^{-i \vec{k} \cdot (\vec{x}-\vec{x}') }
\\ \no
& \times & \, \left[ B_\mu (\vec{x}) \, B_\nu (\vec{x}') \, P(\vec{k}) +
\frac{\delta_{\mu\nu}}{\sqrt{\bar{n}_\mu(\vec{x}) \bar{n}_\nu(\vec{x}')}} \right] \; .
\eea
where the Kronecker delta expresses the absence of shot noise when
cross-correlating different types of objects in the same cell.

If a set of observables $Q_i$ obey a Gaussian distribution with zero mean,
then we can immediately write their Fisher information matrix with respect to
a galaxy survey [\cite{1996ApJ...465...34V,Tegmark:1996bz}]:
\bea
\label{Eq:Fij}
F_{ij} & = & \frac12 {\rm Tr} \, \left( \frac {\partial \log C}{\partial Q_i} 
\, \frac{\partial \log C}{\partial Q_j} \right)
\\ \nonumber
& = & \frac12 {\rm tr} \sum_{\mu\nu\alpha\beta}
\left( C^{-1}_{\mu \nu} \, \frac {\partial C_{\nu \alpha}}{\partial Q_j} \,
C^{-1}_{\alpha \beta} \, \frac{\partial C_{\beta \mu}}{\partial Q_j} \right) \; ,
\eea
where I denote the trace over position- and Fourier-space variables (i.e., the integrals over
those variables) with the lower case. For the sake of clarity, in this Section I will 
leave the integrals over real space and Fourier space explicit.

Suppose that we wish to estimate the bandpowers of the power spectrum, i.e., the 
amplitudes $P(\vec{k}_i)$ on bins $\vec{k}_i$ [for reasons of dimensionality, it is 
more convenient to estimate $\log P(\vec{k}_i)$]. 
The derivatives of the covariance matrix with respect to the bandpowers
are the functional derivatives:
\bea
\no
\frac{\delta C_{\mu \nu} (\vec{x}, \vec{y}) }{\delta \log P(\vec{k}_i)}
&=& B_\mu(\vec{x})  \, 
B_\nu(\vec{y}) \, \int \dk \, e^{- i \vec{k} \cdot (\vec{x} - \vec{y})}
\\ \label{Eq:FunDev}
& \times & P(\vec{k}_i) \, \frac{\delta P(\vec{k})}{\delta P(\vec{k}_i)} 
\\ \nonumber
& = & B_\mu(\vec{x}) \, 
B_\nu(\vec{y}) \, e^{- i \vec{k_i} \cdot (\vec{x} - \vec{y})} \, P(\vec{k}_i) \; ,
\eea
where I have used the fact that, with the conventions used in this paper,
$\delta f(\vec{k})/\delta f(\vec{k}') = (2\pi)^3 \delta_D (\vec{k} - \vec{k}')$.
The Fisher matrix for the power spectrum can then be written as:
\bea
\no
F_P(\vec{k}_i,\vec{k}_j) &=& \frac12 \int d^3 x \, d^3 x' \, d^3 y \, d^3 y' \, 
e^{-i \vec{k}_i \cdot (\vec{x}' - \vec{y}) -i \vec{k}_j \cdot (\vec{y}\,' - \vec{x}) }
\\ \nonumber
& & \times \, \sum_{\mu\nu\alpha\beta}
C^{-1}_{\mu \nu} (\vec{x},\vec{x}') \, B_\nu (\vec{x}') \, P(\vec{k}_i) \, B_\alpha (\vec{y}) \, 
\\ \label{Eq:F2}
& & \times \, C^{-1}_{\alpha \beta} (\vec{y},\vec{y}\,') 
B_\beta (\vec{y}\,') \, P(\vec{k}_j) \, B_\mu (\vec{x}) \; .
\eea

The main problem with Eqs. (\ref{Eq:Fij}) or (\ref{Eq:F2}) 
is that, using position-space pixels, it is not feasible to invert the 
covariance matrix $C_{\mu\nu}$ due to its enormous size.
The formal expression for the covariance does not take us far either, since we 
would then need to solve a system of integral equations:
\be
\label{Eq:IntEqC}
 \int d^3 x' \,  \sum_\nu \, C^{-1}_{\mu \nu} (\vec{x},\vec{x}') \, C_{\nu \alpha} (\vec{x}',\vec{x}'')  
= \delta_{\mu \alpha} \, \delta_D (\vec{x} - \vec{x} '') \; .
\ee

Depending on the concrete case, different approximation schemes can be used 
to invert the covariance matrix.
One such technique consists in consolidating the
the spatial pixels into a much smaller set of pKL eigenfunctions, which drastically 
reduces the dimensionality of the covariance matrix 
[\cite{1998ApJ...499..555T,Tegmark:2003uf,Tegmark:2006az}].
In this case, the particular choice of pKL decomposition 
is justified {\it a posteriori}, in the sense that the particular choice of 
compression is shown to be nearly lossless.

A different scheme that has been used to invert the covariance matrix is 
to try an approximate solution to the formal expression, Eq. (\ref{Eq:COV}) -- 
see, e.g., \cite{Hamilton_97_I}.
The approximation follows from the fact that the average volumetric density $\bar{n}_\mu$ and 
the bias $B_\mu=b_\mu D$ vary slowly as a function of position, compared 
with the exponentials $e^{i \vec{k} \cdot \vec{x}}$ in Eq. (\ref{Eq:COV}).
This means that we can take the integrand of Eq. (\ref{Eq:COV}) and use it
to generate an approximate inverse covariance:
\bea
\label{Eq:InvCo}
C_{\mu\nu}^{-1} (\vec{x},\vec{x}') 
& = & \int \dk \, e^{-i \vec{k} \cdot (\vec{x}-\vec{x}') }
\\ \no
&\times& \left[ B_\mu (\vec{x}) \, B_\nu (\vec{x}') \, P(\vec{k})
+
\frac{\delta_{\mu\nu}}{\sqrt{\bar{n}_\mu(\vec{x}) \bar{n}_\nu(\vec{x}')}} \right]^{-1} \, .
\eea
If we now take the inverse of the expression inside the brackets above to mean
the naive matrix inverse of the $N_t \times N_t$ square matrix, we obtain:
\bea
\label{Eq:InvCo2}
C_{\mu\nu}^{-1}  
& \approx & \int \dk 
\, e^{-i \vec{k} \cdot (\vec{x}-\vec{x}') }
\, \sqrt{\bar{n}_\mu(\vec{x}) \bar{n}_\nu(\vec{x}')}
\\ \no
&\times&
\left[ \delta_{\mu\nu} - 
\frac{\phi_\mu (\vec{x}) \phi_\nu (\vec{x}') \, P(\vec{k})}
{1 + N(\vec{x},\vec{x}') \, P(\vec{k})} 
\right] \, ,
\eea
where $\phi_\mu  = \sqrt{\bar{n}_\mu} \, B_\mu$, and
$N(\vec{x},\vec{x}')=\sum_{\alpha} \phi_\alpha (\vec{x}) \phi_\alpha(\vec{x}')$.
Here $\phi_\mu$ plays the role of a shot noise-corrected 
effective bias for the species $\mu$, in the sense
that the clustering of two species $\mu$ and $\nu$ 
as a function of position (and, therefore, as a function
of redshift), normalized by the underlying matter power spectrum,
has a signal-to-noise ratio proportional to $\phi_\mu \phi_\nu$ -- and this definition
should not be confused with the effective bias as defined in, e.g.,  
\cite{tegmark_3d_2003}.
Substituting the ansatz of Eq. (\ref{Eq:InvCo2}) into Eq. (\ref{Eq:IntEqC}) it can be verified
that the corrections are small when $\phi_\mu$ are smooth functions of the spatial
coordinates  -- in fact, the approximate solution of
Eq. (\ref{Eq:InvCo2}) should be regarded as the first term of a perturbative series, where
the higher-order terms can be obtained by iteration starting with the lowest-order 
solution [\cite{Hamilton_97_I}]. We will show in the next section that this expression 
in fact follows from the use of the classical approximation.

Substituting our ansatz, Eq. (\ref{Eq:InvCo2}), back into Eq. (\ref{Eq:F2}), we obtain:
\bea
\label{Eq:F3}
F_{P, ij} & \approx & 
\frac12 \, P(\vec{k}_i) P(\vec{k}_j) 
 \, \int d^3 x \, d^3 x' \, d^3 y \, d^3 y' \, \dk \dkp
 \\ \nonumber
& \times & 
e^{-i \vec{k}_i \cdot (\vec{x}' - \vec{y}) -i \vec{k}_j \cdot (\vec{y}\,' - \vec{x}) 
- i \vec{k}\cdot(\vec{x}-\vec{x}') - i \vec{k}'\cdot(\vec{y}-\vec{y}\,') }
\\ \nonumber
& \times & 
\frac{N(\vec{x},\vec{x}') + 
P(\vec{k}) \left[ N^2(\vec{x},\vec{x}') - N(\vec{x},\vec{x})N(\vec{x}',\vec{x}') \right]}
{1 + N(\vec{x},\vec{x}') P(\vec{k})}
\\ \nonumber
& \times & 
\frac{N(\vec{y},\vec{y}\,') + 
P(\vec{k}') \left[ N^2(\vec{y},\vec{y}\,') - N(\vec{y},\vec{y})N(\vec{y}\,',\vec{y}\,') \right]}
{1 + N(\vec{y},\vec{y}\,') P(\vec{k}')} \; .
\eea
Integration over $\vec{k}$ and $\vec{k}'$ will select only the positions such that
$\vec{x}'\approx \vec{x}$ and $\vec{y}\,'\approx \vec{y}$, respectively 
(this is the first instance where we need the SP approximation).
Hence, if in Eq. (\ref{Eq:F3}) we make the substitutions 
$N(\vec{x},\vec{x}') \rightarrow N(\vec{x},\vec{x})$,
$N(\vec{x}',\vec{x}') \rightarrow N(\vec{x},\vec{x})$, etc., the terms inside square
brackets cancel, so after integrating over $\vec{x}'$ and $\vec{y}\,'$ we obtain:
\bea
\label{Eq:F4}
F_{P , \, ij} 
& \approx & \frac12 \,
 \int d^3 x \, d^3 y 
 \, e^{ i (\vec{k}_i - \vec{k}_j ) \cdot (\vec{x} - \vec{y}) }
 \\ \no
&\times& \frac{N (\vec{x}) \, P(\vec{k}_i) }{1 + N(\vec{x}) \, P(\vec{k}_i)} \,
\frac{N (\vec{y}) \, P(\vec{k}_j) }{1 + N(\vec{y}) \, P(\vec{k}_j)}  \; .
\eea
Here $ N(\vec{x}) = N(\vec{x},\vec{x})= \sum_\mu \, \bar{n}_\mu \, B_\mu^2 $
plays the role of a total biased effective number density of tracers. 
The integrand of Eq. (\ref{Eq:F4}) is basically the FKP pair window, if we
take the bandpowers $P(\vec{k}_i)$ and $P(\vec{k}_j)$ to be evaluated at 
the same wavenumber, $\vec{k}_i=\vec{k}_j$
-- see, e.g., \cite{Hamilton_97_I}. This equation also shows that (within our
approximations) the best possible 
estimator for the power spectrum has a covariance which is given by 
the inverse of the Fisher matrix above. In fact, the FKP method corresponds 
to weighting pairs by the inverse of the variance of
the power spectrum -- i.e., the weights are the diagonal elements of the Fisher matrix.
An even better estimator is provided by the quadratic method 
[\cite{1998ApJ...499..555T,tegmark_3d_2003}], which employs the full Fisher matrix
-- and therefore takes into account the correlation between estimates of the power
spectrum at different scales.

The advantage of Eq. (\ref{Eq:F4}) is that it splits the problem of
computing the Fisher matrix into two separate Fourier integrals -- in contrast
to the usual approach [\cite{Hamilton_97_I,Hamilton_97_II}].
Nevertheless, I will show below that, at least for the simple case of a survey with a 
top-hat effective number density, $N(\vec{x}) = N_0 \theta(x_0-x)$, 
Eq. (\ref{Eq:F4}) reduces to an expression very similar to that which can be 
obtained directly from the classical limit of the Fisher matrix
[\cite{Hamilton_97_I,Hamilton_97_II}].
In Section \ref{S:QM} I will derive Eq. (\ref{Eq:F4}) using the language and tools of 
quantum mechanics, and in Section \ref{S:A} I will show how to compute the Fisher 
matrix in terms of semi-analytical expressions, in the case of an isotropic 
distribution of galaxies and an isotropic power spectrum.

\vskip 0.3cm

Either by direct computation or by induction from Eq. (\ref{Eq:F4}), 
one can easily write the contributions to the Fisher matrix for
the bandpowers of the power spectrum that come from each one of the 
individual tracers, as well as from their cross-correlations:
\bea
\label{Eq:F5}
F_{P , \, ij}^{\mu \nu}
& \approx & \frac12 \,
 \int d^3 x \, d^3 y 
 \, e^{ i (\vec{k}_i - \vec{k}_j ) \cdot (\vec{x} - \vec{y}) }
 \\ \no
& & \times \, \frac{N_\mu (\vec{x})  \, P(\vec{k}_i) }{1 + N(\vec{x}) \, P(\vec{k}_i)} \,
\frac{N_\nu (\vec{y}) \, P(\vec{k}_i) }{1 + N(\vec{y}) \, P(\vec{k}_j)}  \; ,
\eea
where $N_\mu=\phi_\mu^2$ is the effective biased number density of the tracer species $\mu$.
This expression generalizes the results
of  \cite{Percival:2003pi,White:2008jy,McDonald:2008sh}, 
which were found only in the classical limit and in the SP approximation.
It may be useful to regard the product $N_\mu P \to P_\mu$ as the
power spectrum of each individual species of tracer, and $N \, P=\sum_\mu P_\mu$ 
as the total spectrum -- indeed, that was the notation used in \cite{White:2008jy}.
The Fisher matrices above are additive, as they should be, with the total Fisher matrix,
Eq. (\ref{Eq:F4}), being given by the sum over all species,
$F_{P , ij} = \sum_{\mu\nu} F_{P, ij}^{\mu\nu}$.

Eqs. (\ref{Eq:F4})-(\ref{Eq:F5}) have an intuitive interpretation in terms of the interference
between the information in the phase space cell $(\vec{x},\vec{k}_i)$ and the information 
in the cell $(\vec{y},\vec{k}_j)$. The phase difference 
$e^{-i (\vec{k}_i-\vec{k}_j) \cdot (\vec{x}-\vec{y})}$ can be regarded as a phase space
window function, since it creates a constructive or destructive
interference between the cells which oscillates very rapidly if either
$\vec{k}_i \neq \vec{k}_j$ or if $\vec{x}\neq \vec{y}$. If a pair of galaxies occupies the
same or a nearby spatial cell, $\vec{x} \approx \vec{y}$, then there is no phase difference
and the contribution to the Fisher matrix is basically
flat, with the only sensitivity to the power spectrum $P(k)$ coming from its scale 
dependence -- and in fact, that is the information which is encoded in the effective 
volume $V_{eff}(k)$.
The phase space window function also
takes into account the fact that galaxies separated by a wavelength $\lambda$
contribute most to those wavenumbers which are separated by the harmonics 
of that wavelength,
$ \Delta \vec{k}_n \cdot \Delta \vec{x}=  2 \pi \, n $ ($n=0,1,2, \ldots$) -- but that
contribution falls off with $n$, on account of the faster oscillations of the window function.

The FKP result can now be obtained by taking the functions in phase space to be 
simultaneously localized in position and in Fourier space.
A straightforward way of implementing this approximation is to notice that
$e^{i \Delta \vec{k} \cdot \Delta \vec{x}}$ is in fact a window function 
in phase space which is already normalized to unity over the volume of phase space,
so a stationary phase (SP) approximation naturally leads to the substitution:
\be
\label{Eq:FKPapp}
e^{i \Delta \vec{k} \cdot \Delta \vec{x}}  \rightarrow (2\pi)^3 
\, \delta_D (\Delta \vec{k})
\, \delta_D (\Delta \vec{x}) \; .
\ee
In the SP limit the Fisher matrix for the power spectrum reduces to:
\be
\label{Eq:Fclass}
F_{P , \, ij}
\rightarrow  \frac12 \,
(2\pi)^3 \delta_D(\vec{k}_i-\vec{k}_j) \, \int d^3 x 
\, \left[ \frac{N (\vec{x}) \, P(\vec{k}_i) }{1 + N(\vec{x}) \, P(\vec{k}_i)} \right]^2 \, ,
\ee
which is the familiar result. For the individual species of tracers, the expression is:
\be
\label{Eq:FmnClass}
F_{P , \, ij}^{\mu \nu}
\rightarrow  \frac12 \,
(2\pi)^3 \delta_D(\vec{k}_i-\vec{k}_j) \, \int d^3 x 
\,  \frac{N_\mu (\vec{x})  \, N_\nu (\vec{x}) \, P^2(\vec{k}_i) }
{\left[ 1 + N(\vec{x}) \, P(\vec{k}_i) \right]^2}   \; .
\ee
The integrand in Eq. (\ref{Eq:FmnClass}) is essentially
the result of \cite{Percival:2003pi,White:2008jy}. It is possible
to derive this result also by minimizing the variance of the multi-tracer estimator 
of the power spectrum, as in \cite{Percival:2003pi}, or by directly computing the
covariance of the power spectra between the tracers, as in \cite{White:2008jy}. 
Here I obtained this result from the Fisher matrix in pixel space, which is a 
direct check that the generalization of the FKP pair weighting to many types of 
tracers in fact corresponds to the least-variance estimator.

There are, however, important differences between the more general result, Eq. (\ref{Eq:F5}),
and its SP limit, Eq. (\ref{Eq:FmnClass}): first, the off-diagonal elements of the
Fisher matrix are present in the general expression, but not in the SP limit; and 
second, the manner in which large-scale correlations are manifested in the Fisher 
matrix. To appreciate this difference, consider again
the pathological case already discussed in the Introduction, which I will restate 
here in more generality. Suppose that the tracer species $1$ has
nonzero density at some position $\vec{x}_1$ but it vanishes at the position $\vec{x}_2$;
likewise, the tracer species $2$ has nonzero density at $\vec{x}_2$ but vanishes at 
$\vec{x}_1$. In such a scenario, the integrand of Eq. (\ref{Eq:F5}) for the 
cross-correlation term $F^{12}_{P , \, ij}$ is nonzero when evaluated at those two
points; however, in the SP approximation, Eq. (\ref{Eq:FmnClass}), the integrand
vanishes both at $\vec{x}=\vec{x}_1$ and at $\vec{x}=\vec{x}_2$. Hence, in order to 
fully retain the correlations at different spatial points, we need to keep the distinction 
between the different shells in phase space, which is exactly what the expressions of 
Eqs. (\ref{Eq:F4})-(\ref{Eq:F5}) do.


\section{Fisher matrix in the language of quantum mechanics}
\label{S:QM}

I will now employ some tools borrowed from quantum mechanics in order to derive 
a few useful results, in particular the Fisher matrix for the power spectrum 
$P(\vec{k})$, the Fisher matrix for the effective shot noise-corrected bias, 
$\phi_\mu (\vec{x}) = \sqrt{\bar{n}_\mu} \, B_\mu$, and
the terms of the full Fisher matrix which correlate the estimations of 
$P(\vec{k})$ and $\phi_\mu(\vec{x})$ when
we try to estimate both from the same dataset. This is by no means the only way 
to obtain the full Fisher matrix, but is perhaps the briefest.

As first pointed out by Hamilton [\cite{Hamilton_97_I,hamilton_linear_1997}], the stochastic
variables involved in calculations of the Fisher matrix can be expressed as operators,
while position space and Fourier space are simply two different bases in Hilbert space.
The normalization of the basis vectors (the Dirac bra's and ket's), as well
as the relationship between the position-space basis and the Fourier basis, 
are the usual ones:
\bea
\label{Eq:kk}
\langle \vec{x} | \vec{x}' \rangle &=& \delta_D( \vec{x} -  \vec{x}' ) \, ,
\\
\label{Eq:kkp}
\langle \vec{k} | \vec{k}' \rangle &=& (2\pi)^3 \delta_D( \vec{k} -  \vec{k}' ) \, , 
\\
\label{Eq:xk}
\langle \vec{x} | \vec{k} \rangle &=& e^{- i \vec{k} \cdot \vec{x} } \; .
\eea
In that way, the two-point correlation function and the power spectrum
correspond to the same operator expressed in two different basis,
$\langle \vec{x} | \hat{\xi} | \vec{x}' \rangle = \xi (\vec{x}-\vec{x}')$ and
$\langle \vec{k} | \hat{\xi} | \vec{k}' \rangle = (2\pi)^3 P(\vec{k}) \delta_D(\vec{k}-\vec{k'})$.
By selecting a discrete basis (bins in Fourier space or bins in position space),
operators take the form of matrices.

For simplicity, in this paper I will assume that the average densities $\bar{n}_\mu$ are
directly measured (not estimated), which means that they commute with the derivatives with
respect to any parameters of interest, $\partial \bar{n}_\mu/\partial Q_i = 0$.
In that case, the trace over all indices that defines the Fisher matrix, 
$F_{ij} = (1/2) \, {\rm Tr} \, [ C^{-1} C_{,i} C^{-1} C_{,j} ]$, cancels all the factors
of the average densities that appear outside of the effective biases 
$\phi_\mu = \sqrt{\bar{n}_\mu} \, B_\mu$. This means that we can pull the
number densities out of the covariance matrix, and write 
the Fisher matrix in terms of a renormalized covariance
$F_{ij} = (1/2) \, {\rm Tr} \, [ \hat{C}^{-1} \hat{C}_{,i} \hat{C}^{-1} \hat{C}_{,j} ]$, 
where: 
\be
\label{Eq:RenormCovariance}
\hat{C}_{\mu\nu}(\vec{x},\vec{y}) = \delta_{\mu\nu} + 
\phi_\mu (\vec{x}) \, \xi (\vec{x} - \vec{y}) \, \phi_\nu (\vec{y}) \; .
\ee
Let us then define a renormalized covariance operator as:
\be
\label{Eq:CoOp}
\hat{C}_{\mu\nu} \equiv \delta_{\mu\nu} + \hat{\phi}^\dagger_\mu \, \hat{p}^\dagger \, 
\hat{p} \, \hat{\phi}_\nu \; ,
\ee
where the effective bias operators $\hat{\phi}_\mu=\hat{\phi}^\dagger_\mu$ 
are diagonalized in the position-space basis, 
$ \hat{\phi}_\mu | \vec{x} \rangle = \phi_\mu (\vec{x}) | \vec{x} \rangle$, and
the spectrum operator $\hat{p}=\hat{p}^\dagger$ is diagonalized in the Fourier-space basis,
$ \hat{p} |\vec{k} \rangle = p (\vec{k}) |\vec{k} \rangle$, with $\hat{p}^2=\hat{\xi}$ so that
$ \hat{p}^2 |\vec{k} \rangle =  P (\vec{k}) |\vec{k} \rangle$.
The covariance is clearly hermitian, $\hat{C}_{\mu\nu}^\dagger = \hat{C}_{\nu\mu}$.
Furthermore, taking the expectation value of the covariance operator in
the position-space basis, $\langle \vec{x} | \hat{C}_{\mu\nu} | \vec{x}' \rangle$,
leads to Eq. (\ref{Eq:COV}) -- up to the factors of the average densities, 
which are traced out of the Fisher matrix. The ordering of the operators
in Eq. (\ref{Eq:CoOp}) is in fact unique: if we had defined the renormalized
covariance in any other way, e.g. $\hat{C}_{\mu\nu}^* = 
\delta_{\mu\nu} + \hat{p}^\dagger \hat{\phi}^\dagger_\mu \hat{\phi}_\nu \hat{p}$, 
then its expectation value on a position basis 
$\langle \vec{x} | \hat{C}_{\mu\nu}^* | \vec{x}' \rangle$
would not reduce to the correct expression, Eq. (\ref{Eq:COV}). This is a direct
consequence of the fact that some operators are diagonal in one basis, but not
on the other, which means in particular that the operators $\hat\phi_\mu$ and 
$\hat{p}$ do not commute -- otherwise all possible orderings of the operators 
in the covariance would be equivalent!

In order to invert the (renormalized) covariance matrix operator, it is useful to define the 
{\it total effective covariance} as follows:
\be
\label{Eq:TotCov}
\hat{\cal{C}} \equiv 1 + \sum_\mu   \hat{p} \, \hat{\phi}_\mu \, \hat{\phi}_\mu^\dagger \, \hat{p}^\dagger
= 1 + \hat{p} \, \hat{N} \, \hat{p}^\dagger
\, ,
\ee
where $\hat{N} = \sum_\mu \hat{\phi}_\mu \hat{\phi}_\mu^\dagger$.
In terms of the operator $\hat{\cal{C}}$, the {\em exact} inverse of the covariance operator
is given by the formal expression:
\be
\label{Eq:Inverse}
\hat{C}_{\mu\nu}^{-1} = \delta_{\mu\nu} - 
\hat{\phi}_\mu^\dagger \, \hat{p}^\dagger \, \hat{\cal{C}}^{-1} \, \hat{p} \, \hat{\phi}_\nu \; .
\ee
Of course, this still leaves open the problem of inverting $\hat{\cal{C}}$, but Eq. (\ref{Eq:Inverse}) 
shows that the total effective density $N$ which appears in the definition of $\hat{\cal{C}}$ 
appears quite generically as a result of inverting the covariance of galaxy counts. 
The inverse of $\hat{\cal{C}}$ can be 
obtained either directly from the dataset, or formally in terms of a
perturbative series around the operator $\hat{p} \hat{N} \hat{p}\dagger$ --
in much the same way as was proposed by \cite{Hamilton_97_I}.
This result also shows that, in order to obtain
the exact inverse for the covariance $C_{\mu\nu}$, all that we need is the exact inverse
of the total effective covariance, $\hat{\cal{C}}$, and not the solution to a higher-dimensional 
linear system involving all pairs of every possible type of galaxy. In fact, the higher-dimensional
linear problem has more equations than unknowns, 
and it would be singular were it not for the fact that 
it can be reduced to a single matrix inversion (that of $\hat{\cal{C}}$).

It is instructive to compute the total effective covariance in a mixed basis:
\bea
\label{Eq:CovMix}
\brak \hat{\cal{C}} \ketx &=& \brak (1 + \hat{p} \hat{N} \hat{p}^\dagger) \ketx
\\ \no
&=&  p(\vec{k}) \int \dkp \, \int d^3 y \,
\brak \vec{y} \rangle \langle \vec{y} | \hat{N} \hat{p}^\dagger | \vec{k} ' \rangle 
\langle \vec{k} ' \ketx
\\ \no
& & + \,
e^{ i \vec{k} \cdot \vec{x} }
\\ \no
&=&
\int \dkp e^{ i  \vec{k}' \cdot \vec{x} } 
p(\vec{k}) \tilde{N}(\vec{k} - \vec{k}') p^*(\vec{k}')
\\ \no
& & + \,  e^{ i \vec{k} \cdot \vec{x} }  
 \; ,
\eea
where $\tilde{N}$ is the Fourier transform of the total effective density.
By invoking the classical limit, or equivalently, assuming that 
$N(\vec{x})$ is a smooth function of position, this expression can
be approximated by:
\bea
\label{Eq:CovMix2}
\brak \hat{\cal{C}} \ketx 
\approx
e^{ i \vec{k} \cdot \vec{x} } \left[ 1 +  
\, N(\vec{x}) P(\vec{k}) \right] \; .
\eea

\subsection{Exact Fisher matrix}

In order to derive the Fisher matrix we need to compute the derivatives
$\partial {\cal{C}}_{\mu\nu} /\partial Q_i$, where $Q_i$ are the parameters that we
wish to estimate. In the case where these parameters are the bandpowers
$P_i = P(\vec{k}_i)$, these derivatives can be expressed as the operator:
\be
\label{Eq:DerivP}
\frac{\partial \hat{C}_{\mu\nu}}{\partial P_i}
= \int \dk \, \frac{\partial }{\partial P_i}
\hat{\phi}^\dagger_\mu \, \hat{p}^\dagger \, \ketk \brak \, \hat{p} \, \hat{\phi_\nu}
= \hat{\phi}^\dagger_\mu \, 
| \vec{k}_i \rangle \langle \vec{k}_i | \, \hat{\phi_\nu} \; .
\ee
Using the definition of the total covariance we have that
$ \hat{p} \hat{\phi}_\nu \hat{\phi}_\nu^\dagger \hat{p}^\dagger = \hat{p} \hat{N} \hat{p}^\dagger = \hat{\cal{C}}-1 $, and therefore:
\bea
\label{Eq:dLogC}
\frac{\partial \log \hat{C}_{\mu\alpha}}{\partial \log P_i}
&=& \hat{C}_{\mu\nu}^{-1} \, \frac{\partial \hat{C}_{\nu\alpha}}{\partial \log P_i}
\\ \nonumber
&=& \hat{\phi}^\dagger_\mu \, 
\hat{p}^\dagger \, \hat{\cal{C}}^{-1} | \vec{k}_i \rangle \langle \vec{k}_i |
\hat{p} \, \hat{\phi}_\alpha \; ,
\eea
from which we immediately obtain the Fisher matrix for the (log of the) bandpowers:
\bea
\label{Eq:FishPower}
F_P(\vec{k}_i,\vec{k}_j) 
&=& \frac12 {\rm tr} \left( 
\hat{C}^{-1}_{\mu\nu} \, \frac{\partial \hat{C}_{\nu\alpha}}{\partial \log P_i} \,
\hat{C}^{-1}_{\alpha\beta} \, \frac{\partial \hat{C}_{\beta\mu}}{\partial \log P_j} \right)
\\ \no
&=& \frac12 
\langle \vec{k}_i | (1-\hat{\cal{C}}^{-1}) | \vec{k}_j \rangle
\langle \vec{k}_j | (1-\hat{\cal{C}}^{-1}) | \vec{k}_i \rangle \; .
\eea

Similarly, if we want to estimate the effective biases
$\phi_\alpha(\vec{x})$ from the data, the relevant partial derivatives 
for that Fisher matrix are:
\bea
\no
\frac{\partial \hat{C}_{\mu\nu}}{\partial \phi_\alpha(\vec{x})}
&=&
\frac{\partial }{\partial \phi_\alpha(\vec{x})}
\int d^3 x'  d^3 x'' \, \phi_\mu(\vec{x}') 
| \vec{x}' \rangle \langle \vec{x}'| \hat{p}^\dagger \, 
\\ \no
& & \times \hat{p} \, | \vec{x}'' \rangle \langle \vec{x}''|
\, \hat{\phi}_\nu (\vec{x}'')
\\ \label{Eq:DerivPhi}
&=& \delta_{\mu\alpha} \, \ketx \brax  \, \hat{p}^\dagger  \, \hat{p}  \, \hat{\phi}_\nu^\dagger
\, +  \, \delta_{\nu\alpha}  \, \hat{\phi}_\mu^\dagger  \, \hat{p}^\dagger  \, \hat{p}  \, \ketx \brax  \; .
\eea
A calculation similar to the one performed above for the case of the bandpowers
leads to the Fisher matrix for the effective bias:
\bea
\no
F_{\sigma \gamma} (\vec{x},\vec{y})
&=& \frac12 {\rm tr} \left[ 
\hat{C}^{-1}_{\mu\nu} \, \frac{\partial \hat{C}_{\nu\alpha}}
{\partial \log N_\sigma (\vec{x}) } \,
\hat{C}^{-1}_{\alpha\beta} \, \frac{\partial \hat{C}_{\beta\mu}}
{\partial \log N_\gamma (\vec{y})} \right]
\\ \no
&=& 
\frac14 \, \delta_{\sigma\gamma} \, N_\sigma(x) \, \langle \vec{x} | \vec{y} \rangle 
\langle \vec{y} | p^\dagger \, (1 - \hat{\cal{C}}^{-1} ) \, p \ketx 
\\ \no
& & + \, \frac12 \,  N_\sigma (\vec{x}) N_\gamma (\vec{y}) 
\brax p^\dagger \hat{\cal{C}}^{-1} p | \vec{y} \rangle
\langle \vec{y} | p^\dagger \hat{\cal{C}}^{-1} p \ketx 
\\ \no
& & - \,
\frac18 \, N_\sigma(\vec{x}) N_\gamma (\vec{y}) \,
\left[
\brax p^\dagger p | \vec{y} \rangle
\langle \vec{y} | p^\dagger \hat{\cal{C}}^{-1} p \ketx \right.
\\ \label{Eq:FishBias}
& & + \, \left.
\brax p^\dagger \hat{\cal{C}}^{-1} p | \vec{y} \rangle
\langle \vec{y} | p^\dagger p \ketx 
\right] \; .
\eea
The Fisher matrix for the total effective number density, $N$, can be obtained by
tracing the effective biases,
$F_N = \sum_{\sigma \gamma}  F_{\sigma \gamma} $ .

Finally, we can also compute the cross-terms of the Fisher matrix
that mix the power spectrum estimation with the estimation of the effective bias:
\bea
\no
F_{P \sigma} (\vec{k},\vec{x})
&=& \frac12 {\rm tr} \left[ 
\hat{C}^{-1}_{\mu\nu} \, \frac{\partial \hat{C}_{\nu\alpha}}
{\partial \log P (\vec{k}) } \,
\hat{C}^{-1}_{\alpha\beta} \, \frac{\partial \hat{C}_{\beta\mu}}
{\partial \log N_\sigma (\vec{x})} \right]
\\ \no
&=& 
\frac{N_\sigma(\vec{x})}{4} 
\, \left[ 
\brak p^\dagger \ketx \brax p \, \hat{\cal{C}}^{-1} \ketk \right.
\\ \no
& & + \, \brak \hat{\cal{C}}^{-1} \, p^\dagger \ketx \brax p \ketk
\\ \label{Eq:FishPB}
& & - \, \left. 2 \brak \hat{\cal{C}}^{-1} \,  p^\dagger \ketx \brax p \, \hat{\cal{C}}^{-1} \ketk
\right] \; .
\eea

\subsection{Approximate expressions}

Eqs. (\ref{Eq:FishPower})-(\ref{Eq:FishPB}) are exact, but unless we figure out how
to invert the total covariance ${\hat{\cal{C}}}$, they are not of much use. 
In order to obtain expressions that we can work with, some approximate expression for
that inverse must be produced. The crucial step at this point is to use
the classical approximation, so that $\hat{p}$ commutes with $\hat{\phi}_\mu$, and
therefore operators such as the inverse total covariance can be expressed as
a power series: 
\be
\label{Eq:CommC}
\hat{\cal{C}}^{-1} = (1 + \hat{p} \hat{N} \hat{p}^\dagger)^{-1}
\approx 1 - \hat{p} \hat{p}^\dagger \hat{N} + (\hat{p} \hat{p}^\dagger)^2 \hat{N}^{2} + \ldots \; .
\ee
Just as we used a mixed basis to obtain Eq. (\ref{Eq:CovMix2}),
the matrix elements of the inverse total covariance and other similar operators 
in a mixed basis can be written, in the classical approximation, as:
\bea
\label{Eq:ApCovMix}
\brak \hat{\cal{C}}^{-1} \ketx 
&\approx&
e^{ i \vec{k} \cdot \vec{x} } 
\frac{1}{1 +  \, N(\vec{x}) P(\vec{k}) } \; ,
\\ \label{Eq:Ap1}
\brak \hat{\cal{C}}^{-1} p \ketx 
&\approx&
e^{ i \vec{k} \cdot \vec{x} } 
\frac{P^{1/2}(\vec{k})}{1 +  \, N(\vec{x}) P(\vec{k}) } 
\; ,
\\ \label{Eq:Ap2}
\brak p^\dagger \hat{\cal{C}}^{-1} p \ketx 
&\approx&
e^{ i \vec{k} \cdot \vec{x} } 
\frac{P(\vec{k})}{1 +  \, N(\vec{x}) P(\vec{k}) } 
\; .
\eea
What this means is that:
\bea
\no
\brak \hat{\cal{C}}^{-1} \hat{\cal{C}} \ketkp 
&=& (2\pi)^3 \delta_D( \vec{k} - \vec{k}') 
\\ \no
&=& \int d^3 x \, \brak \hat{\cal{C}}^{-1} \ketx \brax \hat{\cal{C}} \ketkp
\\ \no
&\approx& 
\int d^3 x \, \, e^{ i \vec{k} \cdot \vec{x}} \left[ 1 + N(\vec{x}) P(\vec{k}) \right]^{-1} 
\\ \label{Eq:CovApprox}
& & \times \, e^{  - i \vec{k}' \cdot \vec{x}} \left[ 1 + N(\vec{x}) P(\vec{k}') \right] \; ,
\eea
and a similar expression in the position-space basis.

Using the classical approximation in the Fisher matrix for
the power spectrum, Eq. (\ref{Eq:FishPower}), one readily obtains the same
expression that was derived in the previous Section, Eq. (\ref{Eq:F4}). 
We can also obtain the Fisher matrix for the effective biases in the classical
limit, Eq. (\ref{Eq:FishBias}), in a similar fashion:
\bea
\label{Eq:FB}
F_{\sigma \gamma} (\vec{x},\vec{x}') 
& \approx & 
\frac14 \, 
\delta_{\sigma\gamma} \, \delta_D (\vec{x}-\vec{x}') \,
N_\sigma(\vec{x})
\\ \no
&\times&  \int \dk \, \frac{ N(\vec{x}) P^2(\vec{k})}{1 + N(\vec{x}) P(\vec{k})}
\\ \no
& + & 
\frac18 \,
N_\sigma(\vec{x}) N_\gamma(\vec{x}')
\int \frac{d^3k d^3k'}{(2\pi)^6} e^{ -i (\vec{k}-\vec{k}') \cdot (\vec{x}-\vec{x}')}
\\ \no
&\times&
\, \frac{ P(\vec{k}) \, P(\vec{k}')
\left[ 2 - N(\vec{x}) P(\vec{k}) - N(\vec{x}') P(\vec{k}') \right] }
{[ 1 + N(\vec{x}) P(\vec{k}) ][1 + N(\vec{x}') P(\vec{k}')] } \; .
\eea
The Fisher matrix for the total effective density $N$ can be written as:
\bea
\label{Eq:FN}
F_N (\vec{x},\vec{x}') & \approx & 
\frac18 \, \int \dk \, \dkp \, e^{ -i (\vec{k}-\vec{k}') \cdot (\vec{x}-\vec{x}')}
\\ \no
& & \times \,
\frac{\left[ N(\vec{x}) P(\vec{k}) + N(\vec{x}') P(\vec{k}') \right]^2}
{[1 + N(\vec{x}) P(\vec{k}) ]
[1 + N(\vec{x}') P(\vec{k}') ]} \; .
\eea
Under the classical approximation, the cross-terms of the Fisher matrix which 
mix power spectrum and bias, Eq. (\ref{Eq:FishPB}), become:
\be
\label{Eq:MixFB}
F_{P \sigma} (\vec{k},\vec{x}) \approx \frac12 
\frac{N_\sigma (\vec{x}) \, N (\vec{x}) \,  P^2(\vec{k})}
{[1 + N (\vec{x}) \, P(\vec{k})]^2} \; .
\ee
In terms of the total effective density of tracers, we have:
\be
\label{Eq:MixFN}
F_{P N} (\vec{k},\vec{x}) \approx \frac12 
\left[ \frac{ N (\vec{x}) \,  P(\vec{k})}
{1 + N (\vec{x}) \, P(\vec{k})} \right]^2 \; .
\ee

The main results of this Section can be summarized as follows.
In tandem with the notation of \cite{Hamilton_97_II},
let's define the phase space functions weighting functions (which are 
nothing but the FKP weights):
\bea
\label{Eq:U}
U_\mu(\vec{k},\vec{x}) &=& \frac{ N_\mu(\vec{x}) P(\vec{k})}{ 1 + N(\vec{x}) P(\vec{k})} \; ,
\\ \nonumber
U(\vec{k},\vec{x}) &=& \frac{ N(\vec{x}) P(\vec{k})}{ 1 + N(\vec{x}) P(\vec{k})} = 
\sum_\mu U_\mu(\vec{k},\vec{x}) \; .
\eea
Recall that these weight functions are related to the total covariance operator $\hat{\cal{C}}$,
defined in Eqs. (\ref{Eq:TotCov}) and (\ref{Eq:CovMix}), by 
$\brax (1-\hat{\cal{C}}^{-1} \ketk \approx e^{ - i\vec{k}\cdot \vec{x}} U(\vec{k},\vec{x}) $.
With the help of this function we can write the Fisher matrix for the power spectrum as:
\bea
\no
F_P(\vec{k},\vec{k}') &\approx& \frac12 \int d^3 x \, d^3 x' \, e^{i(\vec{k}-\vec{k}')(\vec{x}-\vec{x}')}
\, U(\vec{k},\vec{x}) \, U(\vec{k}',\vec{x}') 
\\ \label{Eq:GFP}
&=& \frac12 \brak U | \vec{k} ' \rangle \langle \vec{k} ' | U \ketk \; .
\eea
Likewise, the Fisher matrix for the total effective number density 
$N=\sum_\mu \bar{n}_\mu B_\mu^2$ is:
\bea
\label{Eq:GFN}
F_N(\vec{x},\vec{x}') &\approx& \frac18 \, \int \dk \, \dkp \, e^{-i(\vec{k}-\vec{k}')(\vec{x}-\vec{x}')}
\\ \no
& & \times \,
\left\{ 2 U(\vec{k},\vec{x}) \, U(\vec{k}',\vec{x}') \right.
\\ \no
& & + \,  \frac{U^2(\vec{k},\vec{x})}{1-U(\vec{k},\vec{x})}  \left[ 1-U(\vec{k}',\vec{x}') \right]
\\ \no
& & + \left. \left[ 1-U(\vec{k},\vec{x}) \right] \frac{U^2(\vec{k}',\vec{x}')}{1-U(\vec{k}',\vec{x}')} 
\right\} \; . 
\eea
The advantage of these formulas is that they make clear that all
we need to compute are expressions such as:
\bea
\label{Eq:Ukkp} 
\brak U | \vec{k}' \rangle  & = & \int d^3 x \, e^{i (\vec{k}-\vec{k}')\cdot \vec{x}} \, U(\vec{k},\vec{x})
\; ,
\\
\label{Eq:U2kkp} 
\brax \frac{U^2}{1-U} | \vec{x}' \rangle & = &
\int \dk 
\, e^{-i \vec{k}\cdot (\vec{x}-\vec{x}')} \, \frac{U^2(\vec{k},\vec{x})}{1-U(\vec{k},\vec{x})}
\; .
\eea

Last but not least, the cross-terms which mix power spectrum and bias are given by:
\bea
\label{Eq:FMixPN}
F_{P \sigma}(\vec{k},\vec{x})
&\approx& \frac12 U(\vec{k},\vec{x}) U_\sigma(\vec{k},\vec{x}) 
\\ \no
F_{P N}(\vec{k},\vec{x}) &\approx& \frac12 U^2(\vec{k},\vec{x}) =
\sum_\sigma F_{P \sigma}(\vec{k},\vec{x})
\; .
\eea

Equations (\ref{Eq:MixFB})-(\ref{Eq:FMixPN}) therefore express the full Fisher
matrix in the classical approximation, and are the main result of this paper.

\subsection{Generalized FKP formulas}

Finally, let's recover the FKP results by taking the stationary phase 
(SP) limit of Eqs. (\ref{Eq:GFP}) and (\ref{Eq:GFN}).
Recall that, in order to take that limit all we need to do is substitute 
$e^{i \Delta \vec{k} \cdot \Delta \vec{x}} \rightarrow
(2\pi)^3 \delta_D(\Delta \vec{k}) \delta_D (\Delta \vec{x})$.
In that case we obtain:
\bea
\label{Eq:GFPclass}
F_P (\vec{k},\vec{k}') &\rightarrow& 
\frac12 \, (2\pi)^3 \, \delta_D(\vec{k}-\vec{k}') \, \int d^3 x \, U^2(\vec{k},\vec{x}) 
\\ \no
&=& \frac12 \, (2\pi)^3 \, \delta_D(\vec{k}-\vec{k}') V_{eff}(\vec{k}) \; ,
\eea
which is the usual result.

As for the mixed terms of the Fisher matrix, Eqs. (\ref{Eq:MixFB})-(\ref{Eq:MixFN}) or,
equivalently, Eq. (\ref{Eq:FMixPN}), the classical limit result
is already in the SP limit, in the sense used here.

In the case of the Fisher matrix for the effective bias, the result is:
\bea
\label{Eq:GPhiNclass}
F_{\sigma \gamma} (\vec{x},\vec{x}') 
&\rightarrow &
\frac14 \, \delta_D(\vec{x}-\vec{x}') \, \int \dk \, U_\sigma 
\\ \no
& & \times \, \left[ \delta_{\sigma\gamma} \, U \, (1+NP) +  U_\gamma \, (1-NP) \right] \; .
\eea
For the total effective number density, taking either the SP limit of Eq. (\ref{Eq:FN}), or
summing over the species of tracers in the expression above, leads to:
\bea
\label{Eq:GFNclass}
F_N (\vec{x},\vec{x}') &\rightarrow &
\frac12 \, \delta_D(\vec{x}-\vec{x}') \, \int \dk \, U^2(\vec{k},\vec{x}) 
\\ \no
&= & \frac12 \, \delta_D(\vec{x}-\vec{x}') \, \tilde{V}_{eff}(\vec{x}) \; ,
\eea
where the last term on the right-hand side can be interpreted as the 
effective volume in Fourier space. Just as the average number
density of galaxies, the fiducial bias and the growth function affect the
accuracy of the estimations of the bandpowers through $V_{eff}(k)$, the
fiducial power spectrum also affects the accuracy with which we can
measure the bias and the growth function through $\tilde{V}_{eff}(x)$.

An important feature that emerges from the analysis above is that the estimations
of the bandpowers and of the biases are correlated, as shown by 
Eq. (\ref{Eq:FMixPN}). The bottom line is that one should not naively assume 
some normalization for the power spectrum in order to fit a model for the bias,
and then use that bias model in order to fit the power spectrum, while expecting
that the errors in the power spectrum should still be given simply by the Fisher
matrix of Eq. (\ref{Eq:GFPclass}). Because the estimates of the power spectrum are
all correlated with the estimates of the bias, one should in fact
estimate both jointly. The full Fisher matrix derived here allows this joint 
estimation from first principles, which is useful for making more accurate forecasts. 
The expressions above can also be used for a proper treatment
of priors, such as an independent measurement of $\sigma_8$ from either the cosmic
microwave background and/or cluster counts, or 
constraints on bias from gravitational lensing [\cite{2005PhRvD..71d3511S}].
This will be the subject of a forthcoming publication (Abramo 2011, to appear).

\vskip 0.3cm

The results of this Section show a consistent pattern that can be summarized as 
follows. In the same way that we can regard the power spectrum as the density of modes
[and in fact $P(\vec{k})$ has dimensions of a density in Fourier space],
the discussion above implies that we can interpret $N(\vec{x})$ as
the effective density of tracers in position space. 
The combination $\frac12 U^2 = \frac12 N^2  P^2/(1+NP)^2$, in turn, can 
be regarded as
the density of information that can be obtained from a catalog of galaxies
whose biases we don't know, and whose distribution traces some underlying 
matter density whose power spectrum we also don't know. 
This density of information is naturally a phase space object,
since it depends on knowledge about objects which live in Fourier space
and in position space. Hence, the total information contained in
the volume cells $\Delta V_x$ and $\Delta \tilde{V}_k$ is 
$ \frac12 U^2 \Delta V_x \Delta \tilde{V}_k$
-- and this is, in fact, the Fisher information matrix per unit of phase space volume.
The usual FKP Fisher matrix for the power spectrum, evaluated at the 
bin $ \Delta \tilde{V}_k$, is obtained simply by tracing out the position-space volume, 
$F_P (k,k') = \delta_{k,k'} \Delta \tilde{V}_k \int dV_x \frac12 U^2$. The Fisher matrix for the
bias, evaluated at the spatial bin $ \Delta V_x$, is found by tracing out the Fourier space
volume, $F_N(x,x') = \delta_{x,x'} \Delta V_x \int d\tilde{V}_k \frac12 U^2$. And the
elements of the Fisher information matrix that express the correlations
between estimates of the power spectrum and the estimates of bias,
evaluated at the bins $\Delta V_x$ and $\Delta \tilde{V}_k$, is 
$F_{PN} = \frac12 U^2 \Delta V_x \Delta \tilde{V}_k$.

As a curiosity, notice that the properties of $\frac12 U^2$ are very 
similar to those of another object of deep significance in quantum mechanics: 
the Wigner distribution function, which is the phase space 
equivalent of the density matrix [\cite{Wigner32,PeresBook}].
Both the density matrix and the Wigner function can be interpreted as
probability distribution functions -- with the caveat that the Wigner function 
is not necessarily positive, so it is considered a ``quasi-probability'' [\cite{PeresBook}].
The Wigner function has a fundamental role in the physical interpretation
of quantum mechanical states, since it describes how states are 
spread out in Fourier space and in position space.
The Fisher information density $\frac12 U^2$, similarly, describes how 
information is spread over phase space, and what is the probability of 
measuring some parameters in phase space [e.g., the bandpowers
of $P(k)$] within some interval.

\section{Analytical Fisher matrix for a top-hat volume-limited survey}
\label{S:A}

In this Section I will show that, when all variables are isotropic, $P(\vec{k}) \to P(k)$,
$\bar{n}(\vec{x}) \to \bar{n}(x)$ etc., then 
we can express the Fisher matrix for the power
spectrum in the classical limit in terms of analytical formulas. The calculations in the
case of the Fisher matrix for the effective number density are completely
analogous, and the corresponding expressions can be found by exchanging the roles of 
position-space and Fourier-space in the formulas below.

Let's start with the expression for the Fisher matrix for the bandpowers in the 
classical limit, Eq. (\ref{Eq:GFP}), and assume that $P=P(k)$, and $N=N(x)$.
In that case, we can average out the angular dependence of the
Fisher matrix:
\bea
\label{Eq:FPiso}
F_P(k,k) &=&
\int \frac{d^2 \hat{k}}{4\pi} \, \int \frac{d^2 \hat{k}'}{4\pi} \,
F_P(\vec{k},\vec{k}')
\\ \no
&=& 
\frac12
 \int dx \, x^2 
 \int dx' \, x'^2 
\, U(k,x) \, U(k',x')
\\ \no
& & \times \, \int d^2 \hat{k}  \int d^2 \hat{k}' 
\, j_0 (\Delta k \, x) 
\, j_0 (\Delta k \, x')  \; ,
\eea
where hats denote unit vectors, $\hat{k}= \vec{k}/k$, 
 $\Delta k = | \vec{k} - \vec{k} ' |$, 
 and $j_0 (z)=\sin(z)/z$ is the 0-th order spherical Bessel function.
The angular integrals over $\hat{k}$ and $\hat{k'}$ can be performed by writing
$\Delta k = \sqrt{ k^2 + k'^2 - 2 k k' \mu}$, where $\mu=\hat{k} \cdot \hat{k} '$, and
then integrating over $\mu$. At this point, we can use the exact integral:
\bea
\label{Eq:Integral}
I &=& \int_{-1}^1 d\mu \, j_0 (\Delta k \, x) \,  j_0 (\Delta k \, x') 
\\ \no
& = & \frac12 \frac{1}{k \, k' \, x \, x'} 
\left[
{\rm Ci} (\Delta k_+ \Delta x_-) +
{\rm Ci} (\Delta k_- \Delta x_+) \right.
\\ \no
& & \left. -
{\rm Ci} (\Delta k_- \Delta x_-) -
{\rm Ci} (\Delta k_+ \Delta x_+)
\right] \; ,
\eea
where $\Delta k_+ = k+k'$, $\Delta k_- = |k-k'|$, etc., and
${\rm Ci}(z)=-\int_z^\infty dx \cos(x)/x $ is the cosine integral function.
An alternative expression for this integral can be obtained by employing 
Rayleigh's expansion in Eq. (\ref{Eq:GFP}), expanding  each one of the four phases in 
$ \exp[i \vec{k} \cdot (\vec{x} - \vec{x}') + 
i \vec{k}' \cdot (\vec{x}' - \vec{x})] $ into spherical waves, and then integrating over all the 
angles in position space and in Fourier space.
The final result can be recast in terms of the isotropic phase space window 
function ${\cal{W}}$ in two ways:
\bea
\no
{\cal{W}} (k,x;k',x') &=& \frac{1}{2\pi}
\frac{1}{k \, k' \, x \, x'}
\\ \no
& & \times \,
\left[
{\rm Ci} (\Delta k_+ \Delta x_-) +
{\rm Ci} (\Delta k_- \Delta x_+) \right.
\\ \no
& & \left. -
{\rm Ci} (\Delta k_- \Delta x_-) -
{\rm Ci} (\Delta k_+ \Delta x_+)
\right]
\\ \label{Eq:WF}
&=& \frac{2}{\pi} \sum_\ell (2\ell+1)
\\ \no
& & \times \, j_\ell(kx) \, j_\ell(kx') \, j_\ell(k'x) \, j_\ell(k'x') \; .
\eea
With the help of the asymptotic expansion
of the cosine integral function ${\rm Ci}(z) = \gamma + \log z - \frac14 z^2 + {\cal{O}}(z^4) $,
where $\gamma$ is the Euler gamma constant, one can verify 
from the first expression that the window function is regular everywhere, including
the limits
$\Delta k_- \rightarrow 0$ and $\Delta x_- \rightarrow 0$.
From the expression in the second line of Eq. (\ref{Eq:WF}) one can verify that 
the window function is normalized upon
integration over the two-dimensional phase space $(k,x)$, by making use of the identities:
\bea
\label{Eq:BesselDelta}
\int_0^\infty dz \, z^2 \, j_\ell (a z) \, j_\ell (b z) &=& \frac{\pi}{2} a^{-2} \delta_D (a-b) \; ,
\\ \label{Eq:SumBessel}
\sum_\ell (2\ell + 1) j_\ell^2 (z) &=& 1 \; ,
\eea
which then lead immediately to:
\be
\label{Eq:NormW}
\int dk \, k^2 \int dx \, x^2 \, {\cal{W}}(k,x;k',x') = 1 .
\ee
So, it is clear from this expression that the classical limit, in the isotropic case, can be
reached by taking ${\cal{W}}(k,x;k',x') \rightarrow k^{-2} \delta_D(k-k') \, x^{-2}  \delta_D (x-x')$.

In terms of the phase space window function, the Fisher matrices can be written, for
the power spectrum, as:
\bea
\label{Eq:GFPiso}
F_P(k,k') &=& \frac12 (2\pi)^3 \int dx \, x^2 \, \int dx' \, x'^2  \, 
\\ \no
& & \times \, {\cal{W}} (k,x;k',x') \, U(k,x) \, U(k',x') \; ,
\eea
and for the effective number density as:
\bea
\no
F_N(x,x') &=& \frac18 (2\pi)^{-3} \int dk \, k^2 \, \int dk' \, k'^2  \, 
{\cal{W}} (k,x;k',x') \,
\\ \no
& \times &
\left\{ 2 U(k,x) \, U(k',x') + T (k,x) \, [1 - U(k',x')] \right.
\\ \label{Eq:GFNiso}
& & \left. + \, [1-U(k,x)] \, T(k',x') \right\} \; .
\eea

If we use the formula for the phase space window function in terms of the
cosine integrals, then the radial integrations over $x$ and $x'$, or over $k$ and $k'$, are
not separable anymore, as was the case in Eqs. (\ref{Eq:GFP})-(\ref{Eq:GFN}).
However, by writing the phase space window function in terms of the infinite sum over
spherical Bessel functions we can perform the integrations separately.
For the power spectrum Fisher matrix we have:
\bea
\label{Eq:GFPBessel}
F_P(k,k') &=& \frac12 (2\pi)^3 \, \frac{2}{\pi} \, \sum_\ell (2\ell+1) 
\\ \no
& & \times \,
\int dx \, x^2 \, j_\ell(k x) \, j_\ell(k' x) \, U(k,x) \, 
\\ \no
& & \times \,
\int dx' \, x'^2  \, j_\ell(k x') \, j_\ell(k' x') \, U(k',x') \; .
\eea
The two integrals are exactly the same, so for a generic survey all that is needed is
to compute the Hankel transforms of the phase space weighting function $U$:
\be
\label{Eq:Hankel}
U^\ell(k;k') = \sqrt{\frac{2}{\pi}} \int dx \, x^2 \, j_\ell(k x) \, j_\ell(k' x) \, U(k,x) \; .
\ee
The Fisher matrix is then given by the sum:
\bea
\label{Eq:GFPBessel2}
F_P(k,k') &=& \frac12 (2\pi)^3 \, \sum_\ell (2\ell+1) \, U^\ell(k;k') \, U^\ell(k';k) \; .
\eea

In the following Subsection I will show that, for the simple case of a top-hat
density profile, we can employ the dual expressions of the phase space
window function contained in Eq. (\ref{Eq:WF}) in order to obtain an analytical 
solution for the Fisher matrix.

\subsection{Analytical solution: top-hat profile}

\begin{figure}
\begin{center}
\includegraphics[width=8cm]{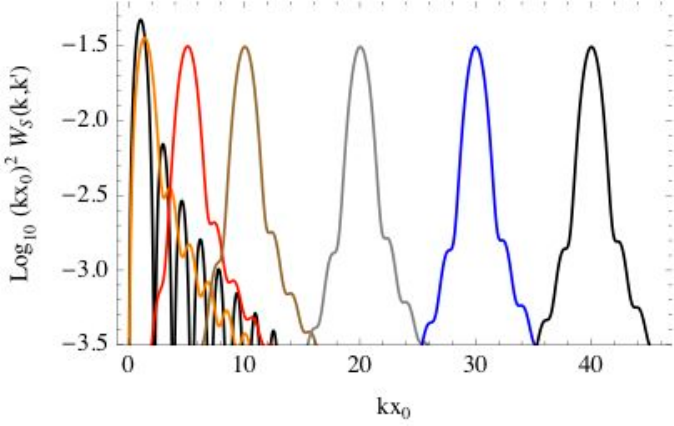}
\caption{Window function $k^2  x_0^2 \, W_{\rm s}^0(k,k')$,
with $k'x_0= 0.1$, 1, 5, 10, 20, 30 and 40 (from left
to right). For $k' x_0 \ll 1$ the window function is essentially independent of
$k'$ for $kx_0 \ll 1$. Notice that even for large $k'$ the window function does not
become more localized around $k=k'$ -- the width of the window function around
the peak is always limited by the size of the survey, $x_0^{-1}$.
}
\end{center}
\label{Fig:wself}
\end{figure}

Now I will show how we can obtain an analytical formula for the Fisher matrix 
$F_P$ in the trivial case of a uniform effective density with a top-hat profile, i.e., 
$N(x) = N_0 \, \theta(x_0-x)$, where $\theta(x)$ is the Heaviside step-function. 
With a top-hat density profile Eq. (\ref{Eq:GFPBessel}) becomes:
\bea
\label{Eq:FPTH}
F_P^0(k,k') &=& 8 \pi^2 \, U_0(k) \, U_0(k') \, \sum_\ell (2\ell+1)
\\ \no
& & \times \,
\int_0^{x_0} dx \, x^2 \, j_\ell(k x) \, j_\ell(k' x) \,
\\ \no
& & \times \,
\int_0^{x_0} dx' \, x'^2  \, j_\ell(k x') \, j_\ell(k' x') \; ,
\eea
where $U_0(k) = N_0 P(k)/[1+N_0 P(k)]$.
The definite integrals above are straightforward, using the identity:
\bea
\label{Eq:IntBessel}
\int_0^1 &dz& z^2 \, j_\ell (az) \, j_\ell (bz) \, 
\\ \no
&=& \frac{1}{a^2 - b^2} 
\left[ b \, j_{\ell-1} (bz) \, j_\ell (az) \, - \, a \, j_{\ell-1} (az) \, j_\ell (bz) \right]
\\ \no
 &=&  \frac{1}{a^2 - b^2} 
\left[ b \, j_{\ell}' (bz) \, j_\ell (az) \, - \, a \, j_{\ell}' (az) \, j_\ell (bz) \right] \; ,
\eea
where, from the second to the third line, I used the recursion relations for
the Bessel functions, \hbox{$j_\ell'(z) = j_{\ell-1}(z) - (\ell-1) j_\ell(z)/z$}.
Using this formula for the integrals over $x$ and $x'$ we obtain that:
\bea
\no
F_P^0(k,k') & = & 8 \pi^2  \,  U_0(k) U_0(k') \frac{x_0^6}{(k^2-k'^2)^2 x_0^4} \, 
 \sum_\ell (2\ell+1) 
 \\ \no
& \times &
\left[ k x_0 \, j_\ell'(k x_0) \, j_\ell(k' x_0) 
- k' x_0 \, j_\ell'(k' x_0) \, j_\ell(k x_0) \right]^2 
\\ \label{Eq:FPTH2}
& \equiv & 8 \pi^2  \, U_0(k) U_0(k') \, x_0^6 \, W_{\rm s}^0(k,k') \; .
\eea
The Fisher matrix above applies only to the self-correlations
of a single species of tracer, and measures the covariance of the auto-correlation
spectrum. For that reason I have called the window function above, $W_{\rm s}^0$, the
self-correlation window function. By taking derivatives of the exact solution for the
full phase space window function, Eq. (\ref{Eq:WF}), it is possible to express
the self-correlation window function in terms of the full isotropic window function:
\bea
\label{Eq:SCWF}
W_{\rm s}^0(k,k') &=& \frac{1}{(k^2 - k'^2)^2 x_0^4} 
\sum_\ell (2\ell+1) 
\\ \no
&\times& 
\left[ k x_0 \, j_\ell'(k x_0) \, j_\ell(k' x_0) 
- k' x_0 \, j_\ell'(k' x_0) \, j_\ell(k x_0) \right]^2 
\\ \no
& = &  \frac{1}{(k^2 - k'^2)^2 x_0^4} \, \frac{\pi}{2} \, 
\left( \frac{\partial}{\partial \log x} \frac{\partial}{\partial \log x'} \right.
\\ \no
& & - 
\left. \left. \frac{\partial}{\partial \log k} \frac{\partial}{\partial \log k'} \right)
\,   {\cal{W}} (k,x;k',x') \right|_{x=x'=x_0} \; .
\eea
Substituting the expression for ${\cal{W}}$ in the first line of Eq. (\ref{Eq:WF}) into
the last expression above we find:
\bea
\label{Eq:SCWF}
W_{\rm s}^0(k,k') &=&   \frac{1}{16} \,
\frac{1}{k k' (k^2 - k'^2)^4 x_0^6} \,
\\ \no
&\times&
\left\{ 8 k k' \left[ k^2 + k'^2 + (k^2 - k'^2)^2 x_0^2 \right] \right.
\\ \no
&  & + \,
(k-k')^4 \cos [ 2 (k+k') x_0 ] 
\\ \no
& & - \, (k+k')^4 \cos [ 2 (k-k') x_0 ] 
\\ \no
&  & + \,
2 (k+k')(k-k')^4 x_0 \sin [ 2 (k+k') x_0 ] 
\\ \no
& & - \left.
2 (k-k') (k+k')^4 x_0 \sin [ 2 (k-k') x_0 ]  \right\} \; .
\eea
A slightly condensed expression can be found using the fact that $z^2 j_{-2}(z) = \cos z + z \sin z$,
which leads to:
\bea
\label{Eq:SCWF2}
W_{\rm s}^0(k,k') &=&   \frac{1}{16} \,
\frac{1}{k k' \Delta k_+^4 \Delta k_-^4 x_0^6} \,
\\ \no
&\times&
\left[ 8 k k' \left( k^2 + k'^2 + \Delta k_+^2 \Delta k_-^2 x_0^2 \right) \right.
\\ \no
&  & + \,
\Delta k_-^4 \, (2 \Delta k_+ x_0)^2 \, j_{-2} (2 \Delta k_+ x_0 ) 
\\ \no
& & -
\left.
\Delta k_+^4 \, (2 \Delta k_- x_0)^2 \, j_{-2} ( 2 \Delta k_- x_0 )  \right] \; .
\eea
In Fig. \ref{Fig:wself} this window function is plotted for some values of $k'$.

The self-correlation window function $W_{\rm s}^0$ is positive everywhere, and is 
highly peaked around $k=k'$ when $k x_0 \gg 1$. However, in contrast to a 
delta-function, it only has support on a finite volume, and therefore it has the
properties that both its width and its maximum height remain finite in the limit $k' \rightarrow k$:
\bea
\no
\displaystyle \lim_{k' \to k} W_{\rm s}^0(k,k') &=&
\frac{1}{256 k^6 x_0^6} 
\left( \cos k x_0 + 4 k x_0 \sin 4 k x_0 \right.
\\ \no
& & \left. - 1 - 8 k^2 x_0^2 + 32 k^4 x_0^4 \right)
\\ \label{Eq:LimSCWF}
& & + {\cal{O}}(k-k') \; .
\eea
In fact, when $k x_0 \gg 1$ the window function can be written as:
\bea
\no
\displaystyle\lim_{k' \to k} \;
\displaystyle\lim_{kx_0 \to \infty} 
W_{\rm s}^0(k,k') &=& \frac{1}{k^2 x_0^2}
\left[ \frac18 - \frac{1}{36} (k-k')^2 x_0^2 \right] 
\\ \label{Eq:Lim2}
& & + {\cal{O}}(k-k')^3 \; .
\eea
This expression shows that even for arbitrarily small scales ($ kx_0 \gg 1$), the finite
size of the survey limits the size of the volume in Fourier space inside which we
can define a bandpower that is linearly independent from the other bandpowers.
The minimal width of bandpowers in the small-scale limit is, from the formula above,
$\Delta k_{\rm min} \sim 3/\sqrt{2} \, x_0^{-1}$.

One can also take the joint limits $k \to k'$ and $k \to 0$, which then result in:
\bea
\label{Eq:Lim3}
\displaystyle \lim_{k' \to k} \;
\displaystyle \lim_{kx_0 \to 0} 
W_{\rm s}^0(k,k') = 
\frac{1}{9}
\left[ 1 - \frac{2}{5} k^2 x_0^2 \right] + {\cal{O}}(k^3) \; .
\eea
This limit shows why the classical approximation is inaccurate 
at large scales (small k): in that regime, the phase space window function is
in fact independent of $k'$ -- i.e., on large scales the Fisher matrix is 
essentially an average over the phase space cells close to the origin. This result
means that the first $k$-bin of a survey has to include {\it all} the modes
$ 0 < k \lesssim \sqrt{2/5} \, x_0^{-1}$. This is, of course, a manifestation of 
cosmic variance, which tells us that no survey can measure structure on scales 
larger than the size of the survey itself.

The preceding discussion implies that the optimal sizes of the bins both in the 
large-scale and in the small-scale regimes are always commensurate with the 
only other scale in the problem, $x_0^{-1}$.
The only exception to this rule would be a spectrum $P(k)$ which has a very sharp
and well-defined feature at some particular scale, such that the spectrum itself 
changes more rapidly than the window function near that scale.

Hence, to summarize the results of this Section, 
we have found that the Fisher matrix for the power spectrum in the case of
a survey with a top-hat number density is given by:
\be
\label{Eq:ExactFP}
F_P^0(k,k')  =  \frac{(2 \pi)^3}{2} \, U_0(k) \, U_0(k') \, \frac{2}{\pi} x_0^6 \, W_{\rm s}^0 (k,k') \; ,
\ee
where $U_0(k) = N_0 P(k)/[1+N_0 P(k)]$ and $W_{\rm s}$ is given by Eq. (\ref{Eq:SCWF}).

Now, I will show that Eq. (\ref{Eq:ExactFP}) is basically identical to the FKP
Fisher matrix that was found, with a slightly different approach, by 
Hamilton [\cite{Hamilton_97_I}]. From that reference, considering only
the lowest-order term in the series for the inverse of the covariance matrix, 
we get that:
\be
\label{Eq:HamiltonF}
F_P^{\rm (FKP)}(k,k')  = \frac12 
\int \frac{d^2 \hat{k}}{4\pi} \, \frac{d^2 \hat{k}'}{4\pi} \, | \tilde{U} (\vec{k}+\vec{k}') |^2 \; ,
\ee
where:
\be
\label{Eq:FTU}
\tilde{U}(\vec{k}) 
= \int d^3 x \, e^{i \vec{k} \cdot \vec{x}} \, U(\bar{k},x) 
= \int d^3 x \, e^{i \vec{k} \cdot \vec{x}} \, \frac{N(x) P(\bar{k})}{1+N(x) P(\bar{k})} 
\; .
\ee
In the expression above, $\bar{k}$ corresponds to a ``trial'' wavenumber that should
be chosen {\it a posteriori} in order to maximize the Fisher matrix (and minimize the 
covariance) for the bandpower that is being estimated. 
The scale $\bar{k}$ is in fact inherited from the inversion of the covariance
matrix, under the approximation that it is diagonal.
It is not entirely clear what sets the correct choice of $\bar{k}$, but it has been common 
practice to take $\bar{k} \to (k + k')/2$ 
[\cite{Hamilton_97_I,Hamilton_97_II,Tegmark_Surveys_1997}].

Under the assumption of isotropy and with a top-hat effective number density
$N(x) = N_0 \theta(x_0-x)$, it is trivial to compute $\tilde{U}$ in the classical limit:
\bea
\label{Eq:ExactU}
\tilde{U}(\bar{k};k) &=& 4 \pi \, U_0(\bar{k}) \, k^{-3} \left[ \sin k x_0 - k x_0 \cos k x_0 \right] 
\\ \no
&=& 4 \pi \, U_0(\bar{k}) \, k^{-3} \, (kx_0)^2 j_1(kx_0)
\; .
\eea
Substituting this expression into Eq. (\ref{Eq:HamiltonF}) we find that the integral is exact, and
the result is in fact:
\be
\label{Eq:CompareFPS}
F_P^{0 \, {\rm (FKP)}} (k,k')  =  \frac{(2 \pi)^3}{2} \, U_0^2(\bar{k})  \, 
\frac{2}{\pi} x_0^6 \, W_{\rm s}^0 (k,k') \; .
\ee
Now compare Eqs. (\ref{Eq:ExactFP}) and (\ref{Eq:CompareFPS}): the phase space window
function is precisely the same in the two expressions, and the only difference
is that the latter equation takes $k=k'=\bar{k}$ in $U_0$. This is a
good approximation only if the $k$ bins are sufficiently large, in which case 
the binned window function is very nearly diagonal.

\subsection{FKP formulas: the stationary phase limit}

Now, let's compare the analytical result of the previous section with the corresponding
FKP formulas (which are in the stationary phase limit). 
Because of the Dirac delta function in the FKP Fisher matrix, it is 
more convenient to compare the averages over bins $k_i$:
\bea
\label{Eq:Fbin}
F_{P,ij}  
&=& \int_{\tilde{V}_i} \dk \, \int_{\tilde{V}_j} \dkp \, F_P(k,k') 
\\ \no
&\approx& 
\frac{\tilde{V}_i}{(2\pi)^3} \, \frac{\tilde{V}_j}{(2\pi)^3} \,
 F_P(k=k_i,k'=k_j) 
\\ \no
& = & \frac{1}{2 (2\pi)^3}
\, U_0(k_i) \, U_0(k_j) \,
\tilde{V}_i \, \tilde{V}_j \, \frac{2}{\pi} \, x_0^6 \, W_{\rm s}^0(k_i,k_j) \; ,
\eea
where $\tilde{V}_i = 4\pi k_i^2 \Delta k_i$ is the volume of the shell in Fourier space
around the $i$-th bin, and in the last expression I assumed that the binning 
is small enough that $F_P$ does not vary too much inside the bin.

The FKP Fisher matrix in the classical limit can be obtained directly from 
Eq. (\ref{Eq:GFPiso}) by taking the SP approximation,
${\cal{W}}(k,x;k',x') \rightarrow k^{-2} \delta_D(k-k') \, x^{-2} \delta_D (x-x')$.
For our top-hat profile, the FKP Fisher matrix for the power spectrum in the
classical limit is:
\be
\label{Eq:FKPTH}
\displaystyle \lim_{\rm class} F_{P,ij}  =
\frac{(2 \pi)^3}{2} \, U_0^2(k_i) \,
\delta_{ij} \, \frac{\tilde{V}_i \, V_0}{(2\pi)^{6}} \; , 
\ee
where $V_0 = 4 \pi \, x_0^3/3$ is the total volume of the survey (in position space, naturally).

It is possible to compare the full expression for the Fisher matrix with its classical
limit, in a way which is completely independent of the phase space weighting function $U_0$ 
-- and, therefore, in a way that does not depend on either the effective density $N_0$
or the fiducial power spectrum $P(k)$.
In fact, all we need to do is to compare the adimensional matrices associated with the
phase space volume:
\be
\label{Eq:Comp}
{\cal{V}}_{ij}^{\rm class} = \tilde{V}_i \; V_0 \; \delta_{ij} 
 \quad \quad v. \quad \quad 
{\cal{V}}^s_{ij} = \tilde{V}_i \; \tilde{V}_j \; \frac{2}{\pi} \; x_0^6 \; W_{\rm s}^0(k_i,k_j) \; ,
\ee

\begin{figure}
\begin{center}
\includegraphics[width=8cm]{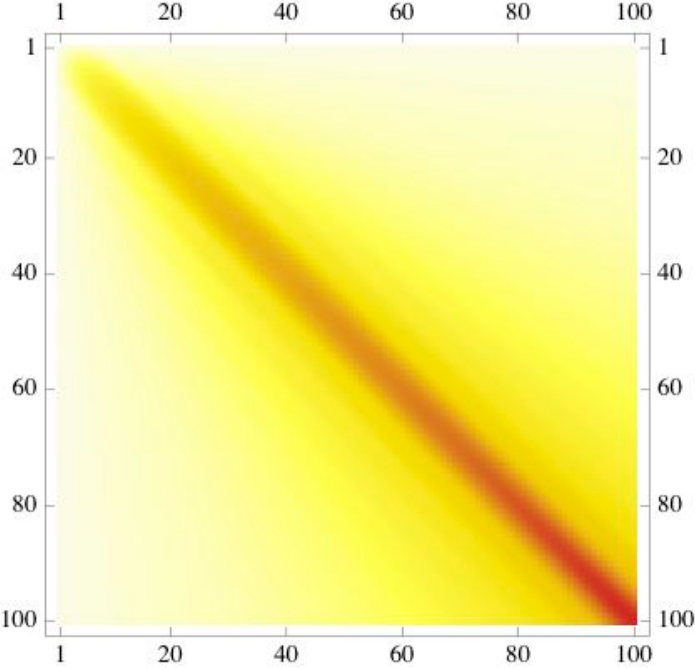}
\end{center}
\caption{
Self-correlation phase space volume matrix ${\cal{V}}^s_{ij}$ 
with 100 equally-spaced bins between $k x_0=0$ and $kx_0=50$.
}
\label{Fig:Vfull}
\end{figure}

\begin{figure}
\begin{center}
\includegraphics[width=8cm]{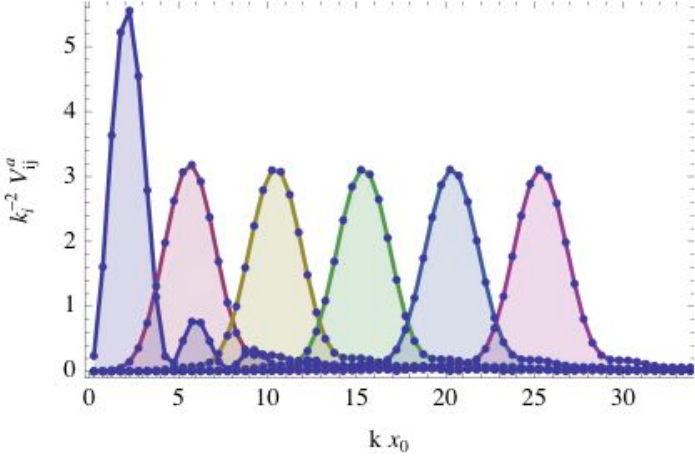}
\end{center}
\caption{
Rows $i$ of the matrix $(k_i x_0)^{-2} {\cal{V}}^s_{ij}$ shown in Fig. \ref{Fig:Vfull}for the bins 
$k x_0=0.5$, 5, 10, 15, 20 and 25 ($j=1$, 6, 11, 16 $\ldots$ 51).
Since $\tilde{V}_i$ is proportional do $k_i^2$, I plot the volume times an 
adimensional pre-factor of $(k_i x_0)^{-2}$.
Above $kx_0 \sim 10$ the rows of the volume matrix 
are essentially self-similar after normalizing for the pre-factor of $(k x_0)^{-2}$.
}
\label{Fig:Vfull2}
\end{figure}

In Fig. \ref{Fig:Vfull} I plot the self-correlation phase space volume 
matrix ${\cal{V}}^s_{ij}$, binned in 100 equally spaced intervals of $\Delta k = 0.5 x_0^{-1}$ 
between $k=0$ and $k x_0 = 50$.
In this 2D representation of the phase space
volume matrix, darker colors denote higher values of phase space volume. 
Obviously, the classical counterpart of this matrix is the diagonal matrix 
${\cal{V}}_{ij}^{class}$. In Fig. \ref{Fig:Vfull2}
I plot some of the rows of the volume matrix, to show how they are spread out over
the $k$ bins. In the classical limit, each curve would be a Dirac delta-function centered on
$k=k'$.

\begin{figure}
\begin{center}
\includegraphics[width=7.5cm]{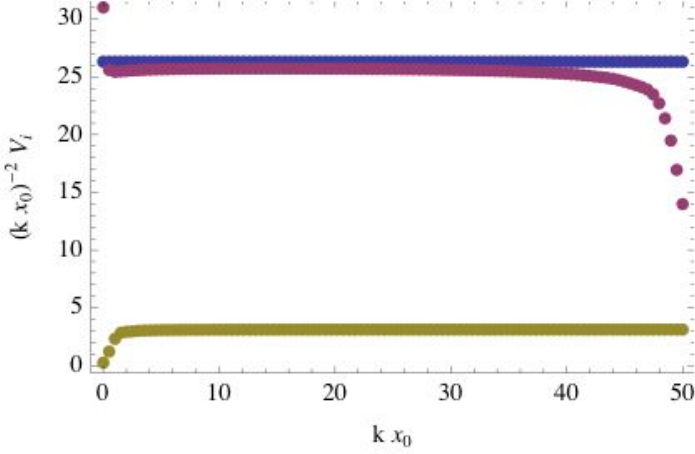}
\end{center}
\caption{
Diagonal elements of the full phase space volume matrix (lower points, yellow in color version),
compared to the same volume in the classical (FKP) approximation (upper points, 
blue in color version). I also show the traces of each individual line 
of the phase space volume matrix (middle points, red in color version), which are
very well approximated by the diagonal elements of the volume matrix in the 
classical approximation, 
simply because the phase space window function is normalized (the end points are off
due to the edges of those bins having been cut-off).
I have employed 100 bins between $k x_0=0$ and $kx_0=50$.
}
\label{Fig:Diag}
\end{figure}

In Fig. \ref{Fig:Diag} I compare the entries of the full phase space volume, ${\cal{V}}_{ij}$, 
with the normalization provided by the classical (FKP) approximation, 
${\cal{V}}_{ij}^{\rm class}$. The upper
set of points (blue in color version) denote the diagonal elements of the phase space
volume in the classical approximation. The lower set of points (yellow in color version) 
denote the diagonal elements of the full phase space volume matrix,
${\cal{V}}_{ii}$. Also plotted are the traces of the rows of the volume matrix, 
${\cal{V}}_i = \sum_j {\cal{V}}_{ij}$ (middle set of points, red in color version).
Since the phase space window function ${\cal{W}}$ is in fact normalized to unity over the
whole volume of phase space, these traces should be equal to their 
classical limit counterparts.
Fig. \ref{Fig:Diag} shows that, indeed, the normalization of the full phase space volume 
matrix is very well approximated by the classical approximation on all but the largest 
scales -- the difference is essentially due to the finite size of the bins. 
The fall-off seen on very small scales (the highest values of $k$) is just an artifact
of cutting off the bins at the edge ($k x_0 = 50$ in our example).

Inspection of Figs. \ref{Fig:Vfull}-\ref{Fig:Vfull2} shows that,
by increasing the size of the $k$ bins, the full Fisher matrix can be well
approximated by the classical limit expression. Under these conditions 
the bandpowers are then approximately uncorrelated -- although
we should always keep in mind that coarse-graining the Fisher matrix leads to loss of 
information, and that this lack of correlation only applies for bins of order 
at least $\Delta k \sim 5 x_0^{-1}$ in our example of a top-hat effective density.
To put that into perspective, for a uniform Hubble-size ($c H_0^{-1} \sim h^{-1} 2997$ Mpc) 
galaxy survey the bins would need to be only as large as 
$\Delta k \sim 3. 10^{-4}$ $h$ Mpc$^{-1}$
for this approximation to be applicable, and for a survey spanning a tenth of a Hubble
volume the $k$ bins would need to be greater than $\Delta k \sim 3. 10^{-3}$ $h$ Mpc$^{-1}$.
As a concrete example, consider the analysis of baryon acoustic oscillations on 
the SDSS-7 performed in \cite{2010MNRAS.401.2148P}: if the volumetric galaxy density of
that dataset were homogeneous and isotropic, 
the minimal size of the bins such that the bandpowers are 
approximately uncorrelated should be of order $\Delta k \sim 0.007$ $h$ Mpc$^{-1}$. 
However, in Figs. 1 and 3 of that paper the bins are spaced only by 
$\Delta k \sim 0.004$ $h$ Mpc$^{-1}$, which means that the datapoints shown in 
those figures are highly correlated. For their statistical analysis, 
\cite{2010MNRAS.401.2148P} fitted cubic
splines on nodes separated by $\Delta k \sim 0.05$ $h$ Mpc$^{-1}$, which gives an
effective bin size of approximately a quarter of that separation, i.e., 
$\Delta k_{eff} \sim 0.0125$ $h$ Mpc$^{-1}$, which is close to the limit I computed
above assuming a top-hat density profile of galaxies. Using a more realistic
distribution of galaxies as a function of redshift and angular position in the sky 
would only make this problem worse.

\begin{figure}
\begin{center}
\includegraphics[width=8cm]{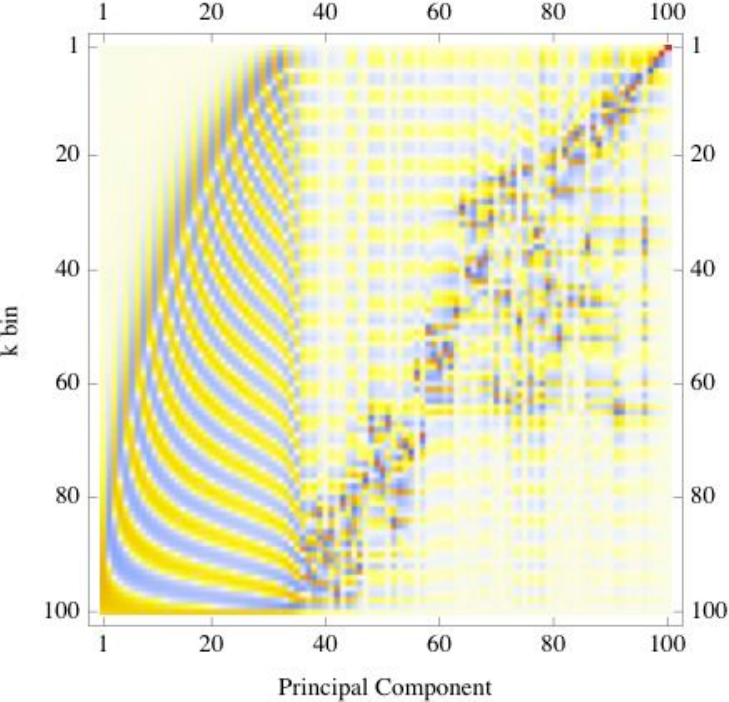}
\end{center}
\caption{
Principal components of the Fisher matrix, binned in 100 equally
spaced intervals between $kx_0=0$ and $k x_0=50$. The horizontal axis corresponds
to each principal component, ranked by their eigenvalues, and the vertical axis 
(from top to bottom) corresponds to the $k$ bins.
}
\label{Fig:PCA}
\end{figure}

\begin{figure}
\begin{center}
\includegraphics[width=7cm]{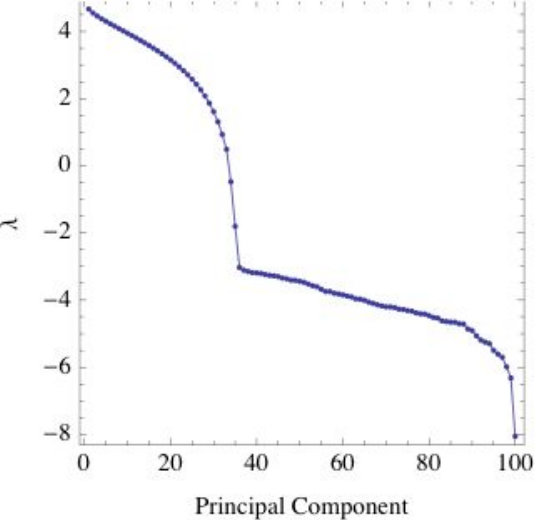}
\end{center}
\caption{
Eigenvalues of the principal components of the Fisher matrix.
It is clear from both Fig. \ref{Fig:PCA} and \ref{Fig:PCA2} 
that there are only about 35 principal values with non-negligible
eigenvalues, which means that only as many linearly independent bandpowers can
be estimated from the data. The plots also show that, among the non-negligible
principal components, the highest-ranked correlate with large values of
$k$, and the lowest-ranked involve essentially the small-$k$ bins.
}
\label{Fig:PCA2}
\end{figure}

As mentioned above, one should keep in mind that by increasing the size of the bin 
we lose some amount of information by washing out the Fisher matrix. 
On the other hand, for practical and numerical purposes
it is inefficient to keep an excessively large number of bins.
In that sense, it is interesting to examine how many linearly independent modes 
are encoded in the Fisher matrix, by making a principal component analysis (PCA) on it.
The result of the PCA decomposition fn our Fisher matrix with 100 equally 
spaced bins is presented in Fig. \ref{Fig:PCA}, where the principal values are
plotted as a function of $k$ for the 100 bins. Fig. \ref{Fig:PCA2} shows
the eigenvalues corresponding to the principal values. From Fig. \ref{Fig:PCA2} it can be 
seen that only about 35 components are relevant for this Fisher matrix, and from 
Fig. \ref{Fig:PCA} we see that the highest-ranked ones probe the small scales (large $k$),
whereas the lowest-ranked amongst the 35 non-trivial principal components span the
large scales (small $k$). The 65 lowest-ranking principal 
values have negligible eigenvalues, and it is clear that they carry no information 
whatsoever. This is an indication that the optimal average size of the bins in the 
case of a top-hat survey should be at most of the order of $\Delta k \sim 50 \, 
x_0^{-1}/35 \simeq 1.4 \, x_0^{-1}$. However, in that case one should be careful to 
include the cross-correlations between the different bandpowers, as not doing so 
would lead to an overestimation of the constraints.

\subsection{Analytical solution for two species of tracers 
with top-hat density profiles, including the cross-correlations}

Now I will generalize the results of the previous section to the case where we have
two species of tracers. 
The main distinction with the previous sections is that now the total number 
density is the sum of two different top-hats,
$N(x) = N_1(x) + N_2(x)$, where $N_\mu(x) = N_\mu \, \theta (x_\mu - x)$.
Here I will assume that the survey of species 1 is dense but shallow, and the
survey of species 2 is sparse but deep, so $N_1 > N_2$ and $x_1 < x_2$.
This would be the case, e.g., of a homogeneous survey of luminous red galaxies 
limited to $z \lesssim 1$, and a survey of quasars or Ly-$\alpha$ absorption 
systems limited to $z \lesssim 3$.

With this in mind, we can write the phase space weighting function
$U_1$ of the previous Section as:
\bea
\label{Eq:U1}
U_1 (k,x) &=& \frac{ N_1 \theta (x_1 - x)P(k)}{1 + [N_1 \theta (x_1-x) + N_2 \theta (x_2-x)]P(k)}
\\ \no
& = & \frac{ N_1 P(k)}{1 + N_{12} P(k)} \theta (x_1-x) \; ,
\eea
where $N_{12}=N_1 + N_2$ is the total effective number density for $0 < x \leq x_1$.
For the tracer species 2 the weighting function is expressed as:
\bea
\label{Eq:U2}
U_2 (k,x) &=& \frac{ N_2 \theta (x_2 - x) P(k)}{1 + [N_1 \theta (x_1-x) + N_2 \theta (x_2-x)] P(k)}
\\ \no
& = & \left[ \frac{ N_2 P(k)}{1 + N_{12} P(k)} - \frac{ N_2 P(k)}{1 + N_2 P(k)} \right] 
\theta (x_1-x) 
\\ \no
& & + \frac{ N_2 P(k)}{1 + N_2 P(k)} \theta (x_2-x) ; .
\eea
Clearly, then, we can redefine the two effective densities so that they reflect
the two distinct top-hats. Collecting the amplitudes of each individual
top-hat profile, we obtain:
\bea
\label{Eq:U1p}
U_{1}' (k,x) & = & \left[ \frac{ N_{12} P(k)}{1 + N_{12} P(k)} - \frac{ N_2 P(k)}{1 + N_2 P(k)} \right] 
\, \theta (x_1-x) 
\\ \no
&\equiv& U_{1}' (k) \, \theta (x_1-x) \; ,
\\ 
\label{Eq:U2p}
U_{2}' (k,x) & = & \frac{ N_2 P(k)}{1 + N_2 P(k)} \, \theta (x_2-x) 
\\ \no
&\equiv& U_{2}' (k) \, \theta (x_2-x) \; .
\eea

Therefore, when computing the total Fisher matrix for two species
of tracers one should include the two self-correlation Fisher matrices,
as discussed in the previous section, using either $U_{1}'(k)$ for the
self-correlation of species 1, or $U_{2}'(k)$ for the self-correlation
of species 2. In addition, we must also include the Fisher matrix for
the cross-correlation between the two effective top-hats of 
Eqs. (\ref{Eq:U1p})-(\ref{Eq:U2p}), which I discuss now.

\begin{figure}
\begin{center}
\includegraphics[width=7.5cm]{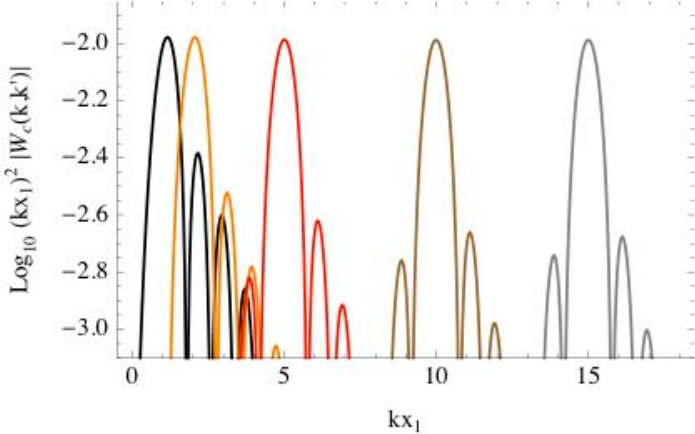}
\end{center}
\caption{Cross-correlation phase space window function $k^2  x_1^2 \, W_{\rm c}^{12}(k,k')$,
with $x_2=2 x_1$. The modes plotted are, from left to right, 
$k' x_1= 1$, 2, 5, 10, and 15. As was the case for the self-correlation window function,
for $k' x_1 \ll 1$ the window function is essentially independent of
$k'$. The most important difference between the two window functions is that the cross
window function can become negative -- notice that I have plotted the log of the absolute value
of $k^2 W_{\rm c}$, but the positive values are always found at the peak of the window function,
and the spikes to negative infinity mark the transition from positive to negative values or 
vice-versa. The width of the window function around
the peak is given by the inverse of the largest length scale of the survey
(in this example, $x_2$).
}
\label{Fig:wcross}
\end{figure}

The Fisher matrix for the cross-spectrum arises from the 
cross-correlation between two species of tracers of large-scale structure. 
From Eq. (\ref{Eq:GFPBessel}), and according to the discussion of Section \ref{S:F},
the Fisher matrix for the cross-correlation between the top-hat of radius $x_1$ and 
the top-hat density of radius $x_2$ is given by:
\bea
\label{Eq:FPTHcross}
F^{12}_P(k,k') &=& 8 \pi^2 \, U_{1}'(k) \, U_{2}'(k') \, \sum_\ell (2\ell+1)
\\ \no
& & \times \int_0^{x_1} dx \, x^2 \, j_\ell(k x) \, j_\ell(k' x) \,
\\ \no
& & \times \int_0^{x_2} dx' \, x'^2  \, j_\ell(k x') \, j_\ell(k' x') \; .
\eea
Using the same derivation that was used to arrive at Eq. (\ref{Eq:FPTH2}), I get:
\bea
\label{Eq:SCWFcross}
W_{\rm c}^{1\, 2}(k,k') &=& \frac{1}{(k^2 - k'^2)^2 \, x_1^2 \, x_2^2} 
\\ \no
& & \times \sum_\ell (2\ell+1) \, g_\ell^1 (k,k') \, g_\ell^2 (k,k') \; ,
\eea
where:
\bea
\label{Eq:Defg}
g^{i}_\ell (k,k') = k \, x_i \, j_\ell'(k x_i) \, j_\ell(k' x_i) - k' x_i \, j_\ell'(k' x_i) \, j_\ell(k x_i).
\eea
Since $g_\ell^i (k,k')$ is of order ${\cal{O}}(k-k')$ in the limit
$k' \to k$, the cross window function is well-behaved everywhere.
The most important distinction between the self-correlation window function 
$W_{\rm s}$ and the cross-correlation one, $W_{\rm c}$, is that the former is always
positive, whereas the latter is positive at its peak (at $k'=k$), but presents damped
oscillations between positive and negative values away from the peak. 
Another important difference between the self-correlation and the cross-correlation
window functions is that I was able to find an analytical expression for the 
former, but not for the latter, which is left in the form of the infinite sum,
Eq. (\ref{Eq:SCWFcross}). The situation is not as dire as it may seem, since the spherical
Bessel functions $j_\ell(z)$ are highly peaked around $z\sim \ell$, and Limber-type
approximations allow us to cut off the infinite sum to a small number of terms
with minimal loss of precision.
In Fig. \ref{Fig:wcross} I plot some of the modes of the cross window function in the case 
$x_2=2 x_1$.

The full Fisher matrix for the power spectrum in the case of two top-hat profiles is 
therefore given by the combination of the self-correlation and the cross-correlation 
terms corresponding to the two top-hats above, with amplitudes $U_1'(k)$ and
$U_2'(k)$. The explicit expression is:
\bea
\label{Eq:FPall}
F_P(k,k') & = &  
8 \pi^2 \, \left\{  x_1^6 \, U_{1}'(k) \, U_{1}'(k') \, W_{\rm s}^1 (k,k') \right.
\\ \no
& & +
 x_2^6 \,  U_{2}'(k) \, U_{2}'(k') \,  W_{\rm s}^2 (k,k') 
\\ \no
& & +
\, x_1^3 x_2^3 \, \left[ U_{1}'(k) \, U_{2}'(k') \, W_{\rm c}^{12} (k,k') \right.
\\ \no
& & \left. \left. + U_{2}'(k) \, U_{1}'(k') \,  W_{\rm c}^{21} (k,k') \right]  \right\}
\; ,
\eea
and the binned Fisher matrix is, therefore:
\bea
\label{Eq:FPallbin}
F_{P,ij} & = &  \frac{1}{2(2 \pi)^3} 
\, \left\{ U_{1,i}' \, U_{1,j}' \, {\cal{V}}_{ij}^{s \, (1)}
+
\, U_{2,i}' \, U_{2,j}' \, {\cal{V}}_{ij}^{s \, (2)} \right.
\\ \no
& & \left. +
\, \left[ U_{1,i}' \, U_{2,j}' + U_{2,i}' \, U_{1,j}' \right] \,  {\cal{V}}_{ij}^{c}  \right\}
\; ,
\eea
where ${\cal{V}}^c$ is defined in terms of the cross-correlation window function $W_{\rm c}$ 
in the same way as ${\cal{V}}^s$ was defined in terms of $W_{\rm s}$ in Eq. (\ref{Eq:Comp}):
\be
\label{Eq:DefVc}
{\cal{V}}_{ij}^c = \tilde{V}_i \, \tilde{V}_j \, \frac{2}{\pi} \, x_1^3 \, x_2^3 \, W_{\rm c}(k_i,k_j) \; .
\ee

The SP approximation to the full Fisher matrix can be reached directly 
from Eq. (\ref{Eq:GFPiso})
by taking the limit ${\cal{W}} (k,x;k',x') \to k^{-2} \delta_D(k-k') \, x^{-2}  \delta_D (x-x')$,
and then integrating over the volumes of the $k$ bins:
\bea
\label{Eq:FKP12}
F_{P,ij}^{\rm FKP} &=& 
\frac{1}{2 (2\pi)^3} \, \tilde{V}_i \, \delta_{ij} 
\\ \no
& & \times
\left[ \frac{4\pi}{3} \, x_1^3 \, U_{12,i}^2 + \frac{4 \pi}{3} \, (x_2^3 - x_1^3) \, U_{2,i}'^2 \right] \; ,
\eea
where $U_{12,i} = N_{12} P(k_i)/[1 +N_{12} P(k_i)] = U_1' + U_2'$.
The relevant difference with respect to the analytical expression in the classical limit is 
that, because of the SP approximation, all the signal from the cross-correlation 
between 1 and 2 is already implicitly included in the amplitudes $U_{12}$ and $U_2'$.
This becomes clearer if we rewrite the SP limit of the full Fisher matrix as:
\bea
\label{Eq:FKP12prime}
F_{P,ij}^{\rm FKP} 
&=& 
\frac{1}{2 (2\pi)^3} \, \tilde{V}_i \, \delta_{ij} \, \frac{4\pi}{3}
\\ \no
& & \times
\left[ x_1^3 \, U_{1,i}'^2  +  x_2^3 U_{2,i}'^2 
+ x_1^3 \left( U_{12,i}^2 - U_{1,i}'^2 - U_{2,i}'^2 \right) \right] 
\\ \no
&=& 
\frac{1}{2 (2\pi)^3} \, \tilde{V}_i \, \delta_{ij} \, \frac{4\pi}{3}
\\ \no
& & \times
\left[ x_1^3 \, U_{1,i}'^2  +  x_2^3 \, U_{2,i}'^2 
+ 2 \, x_1^3 \, U_{1,i}' \, U_{2,i}'
\right] 
\; .
\eea
Compare now this last expression with Eq. (\ref{Eq:FPallbin}). The
self-correlation term for species 1 has already been analyzed in the previous sections,
and the self-correlation term for species 2 is precisely the same, except for the
scaling $k \to k \times x_1/x_2$.
The comparable cross-correlation terms are $\tilde{V}_i V_1 \delta_{ij}$
and ${\cal{V}}_{ij}^c$, as defined in Eq. (\ref{Eq:DefVc}).

\begin{figure}
\begin{center}
\includegraphics[width=8cm]{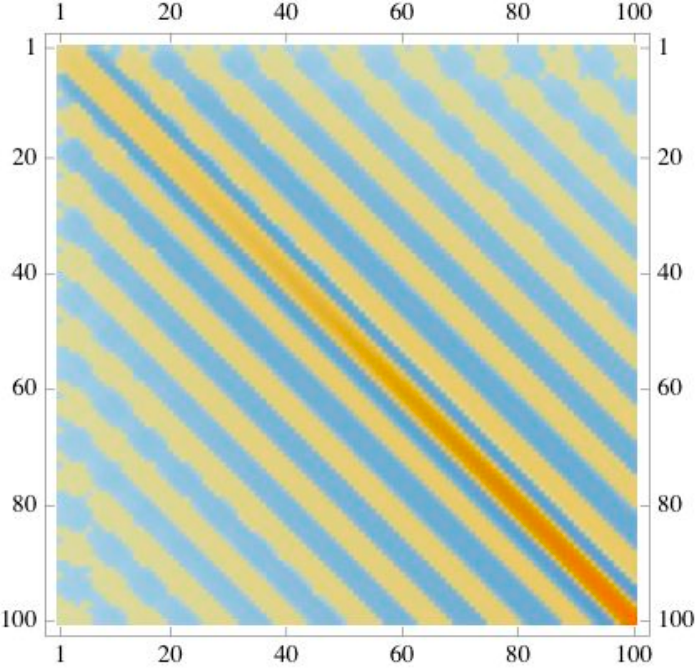}
\end{center}
\caption{
Cross-correlation phase space volume matrix ${\cal{V}}^c_{ij}$, 
with 100 equally-spaced bins between $k x_1=0$ and $kx_1=50$,
and assuming $x_2=2 x_1$.
}
\label{Fig:Vcross}
\end{figure}

\begin{figure}
\begin{center}
\includegraphics[width=6.5cm]{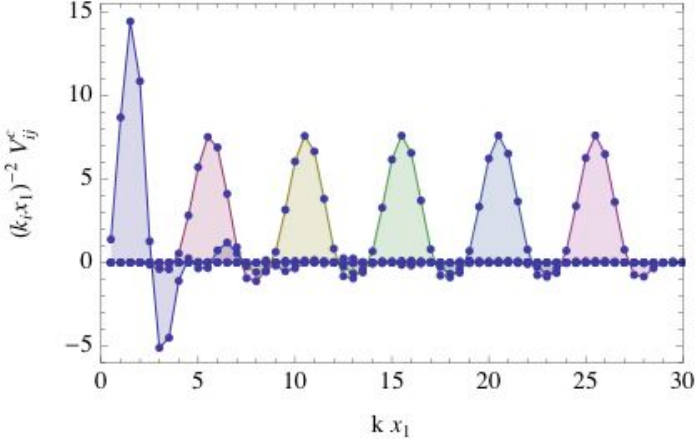}
\end{center}
\caption{
Rows of $k_i^2 \, {\cal{V}}^c_{ij}$ for the bins $k x_1=0.5$, 5, 10, 15, 20 and
25 (from left to right, respectively).
}
\label{Fig:Vcross2}
\end{figure}

I show the cross-correlation phase space volume in Figs. \ref{Fig:Vcross} and \ref{Fig:Vcross2}.
On Fig. \ref{Fig:Vcross}, the rows of the volume matrix are plotted for the
first few $k$ bins. As discussed above, the phase space window function 
(and, therefore, the volume matrix) can be negative for the cross-correlation term.
Comparing Figs. \ref{Fig:Vfull2} and \ref{Fig:Vcross2} we see that the cross-correlation
volume matrix is narrower around the diagonal. However, this is a simple
consequence of the inclusion of the second species of tracer, whose top-hat
profile has a radius $x_2 = 2 x_1$. Naturally, the width of the cross-correlation
volume matrix in $k$ bins is dominated by the inverse of the largest scale, 
which in this example is $\Delta k \sim x_2^{-1} = 0.5 x_1^{-1}$ .

\begin{figure}
\begin{center}
\includegraphics[width=7.5cm]{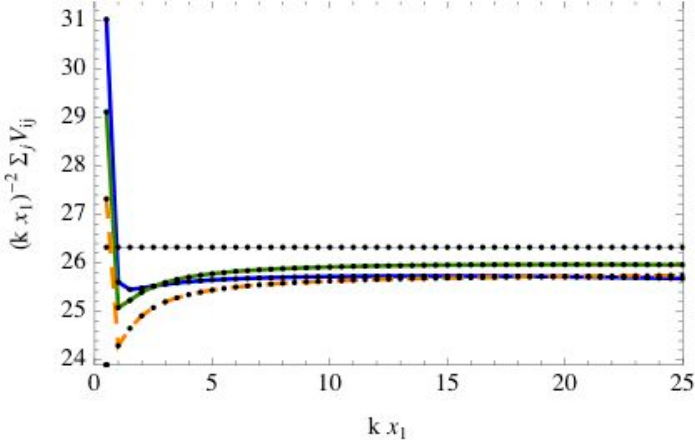}
\vskip 0.5cm
\includegraphics[width=7.5cm]{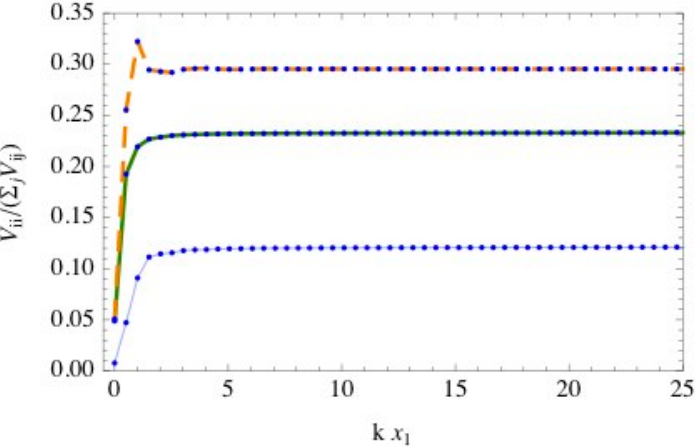}
\end{center}
\caption{
Upper panel: the horizontal row of points are the classical and SP approximations
to the phase space volume matrix for the cross-correlation between tracer species 1 and 2.
The lower and upper solid lines (blue and green in color version) correspond to the
trace of the rows of the self-correlation phase space matrices of species 1 and 2, respectively
-- i.e., $\sum_j {\cal{V}}_{ij}^{s (1,2)}$.
The dashed line (orange in color version) corresponds to the trace of the cross-correlation
phase space matrix.
Lower panel: ratios of the diagonal elements of the phase space volume matrices
to their traces, ${\cal{V}}_{ii}/\sum_j {\cal{V}}_{ij}$. From the lower to the upper lines, 
self-correlation of species 1 (lower solid line, blue in color version), 
self-correlation of species 2 (middle solid line, green in color version), 
and cross-correlation ${\cal{V}}_{ii}^{c}$ (upper dashed line, orange in color version).
I employ 100 bins between $k x_1=0$ and $kx_1=50$.
}
\label{Fig:Diagcross}
\end{figure}

In Fig. \ref{Fig:Diagcross} I compare the phase space volume with its SP 
approximation, for both the self-correlation terms and for the cross-correlation as well. 
In order to compare the volume matrices on
an equal footing, I have normalized the volume of species 2 to the volume of species 1.
In the upper panel I have plotted $\tilde{V}_i V_1$, as well as the traces of
each row of the volume matrices, $\sum_j {\cal{V}}_{ij}^{s(1)}$,
$V_1/V_2 \times \sum_j {\cal{V}}_{ij}^{s(2)}$ and $ \times \sum_j {\cal{V}}_{ij}^{c}$.
From the upper panel we can verify that there is good agreement between the
SP approximation and the analytical result in the classical approximation 
at intermediate scales, where the analytical result is only
$\sim$ 2-3\% below the SP approximation. This means that, by taking large enough 
$k$-bins one can recover the FKP result for the Fisher matrix from the result in
the classical approximation.
Only at the very largest scales (the single bin between $0 < k x_1 \leq 0.5$) 
the SP approximation fails, and 
the FKP Fisher matrix understates the constraining power of the survey.

The lower panel of Fig. \ref{Fig:Diagcross} 
shows the ratios of the diagonals of the phase space volume matrices
to the traces of their respective rows -- which is a measure of how diagonal those matrices 
are. Tracer species 1, which occupies the smallest volume, is the least diagonal, and
tracer species 2, which spans length scales roughly double those of species 1, has
a more diagonal Fisher matrix, by a factor of two, approximately. The
cross-correlation Fisher matrix, in fact, appears to be the most diagonal of the Fisher 
matrices. However, that is partly an artifact coming from the wings of the cross-correlation 
window function, which are negative  -- see Fig2. \ref{Fig:Vcross}-\ref{Fig:Vcross2}. 
When we account for this (by, e.g.,
using the squares of the window functions as the normalization), the cross-correlation 
Fisher matrix comes out to be approximately as diagonal as the self-correlation
Fisher matrix of the tracer species with the largest volume -- in our case, species 2.

\begin{figure}
\begin{center}
\includegraphics[width=8cm]{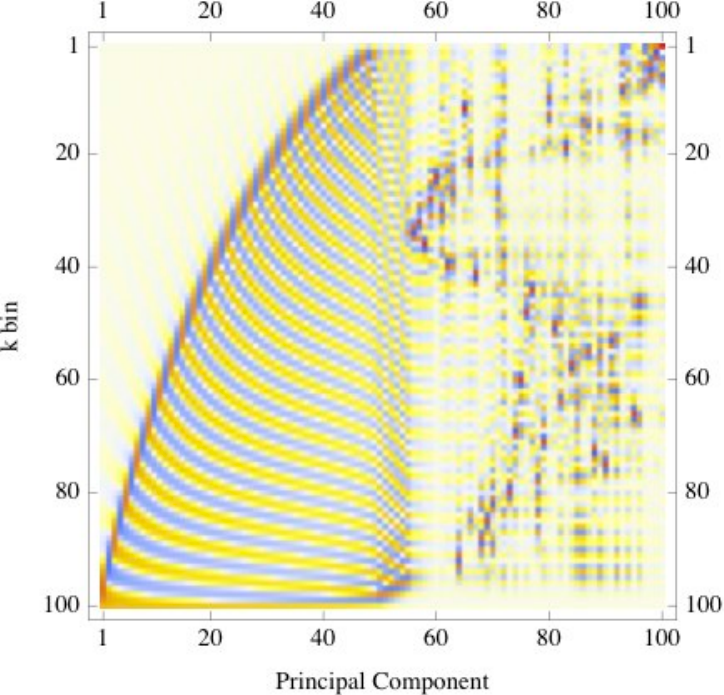}
\end{center}
\caption{
Principal components of the cross-correlation Fisher matrix for the
power spectrum, binned in 
100 equally spaced intervals between $kx_0=0$ and $k x_0=50$. The horizontal 
axis corresponds to each principal component, ranked by their eigenvalues, and 
the vertical axis (from top to bottom) corresponds to the $k$ bins.
}
\label{Fig:PCA_cross}
\end{figure}

\begin{figure}
\begin{center}
\includegraphics[width=8cm]{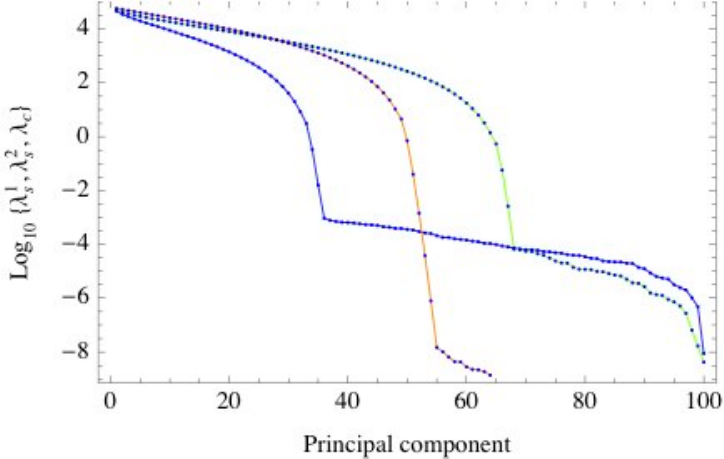}
\end{center}
\caption{
Eigenvalues of the principal components of the Fisher matrices for
the self-correlation of tracer species 1 with itself (left line and points, blue in color version),
for the self-correlation of species 2 with itself (right line, green in color version),
and for the cross-correlation between the two species (middle line, orange in color version).
}
\label{Fig:PCA_cross2}
\end{figure}

It is also useful to perform a PCA analysis on the cross-correlation phase space
volume, as was done for the self-correlation volume in the previous section.
On Fig. \ref{Fig:PCA_cross} I show the principal components of ${\cal{V}}^c$
as a function of the $k$ bins. 
On Fig. \ref{Fig:PCA_cross2} I show the eigenvalues of the
principal values for the self-correlations of species 1 and 2, as well as the cross-correlation
between the two species.

The results of this Section can be easily generalized to an arbitrary 
number of species of tracers, and to any (isotropic) number density -- not
only uniform densities. 



\section{Discussion}

In this paper I have shown how to compute the Fisher matrix for galaxy surveys, including
the cross-correlations between different cells in position space and in Fourier space.
In the stationary phase approximation these cross-correlations are discarded --
Hamilton [\cite{Hamilton_97_I,hamilton_linear_1997}] refers to this case as the 
``classical limit''. 

I have also shown how to obtain the Fisher matrix for multiple species of tracers
of large-scale structure from the covariance of counts of galaxies in cells. 
The final formulas, after taking the classical and SP limits, generalizes the results
previously obtained by \cite{Percival:2003pi,White:2008jy,McDonald:2008sh}. 
However, I have also
obtained the Fisher matrix using only the classical limit, which solves some 
(probably minor) inconsistencies of those formulas. I have also shown that,
in order to invert the covariance matrix for the full dataset with all species of 
tracers, all that is needed is the inversion of a single matrix (or operator), 
$\hat{\cal{C}}$, and not the inversion of a large set of linear 
equations -- see Eq. (\ref{Eq:TotCov}) and the following discussion.
The main results are summarized by
Eqs. (\ref{Eq:U})-(\ref{Eq:FMixPN}), and their classical limits are shown in Eqs. 
(\ref{Eq:GFPclass})-(\ref{Eq:GFNclass}).

The full Fisher matrix in the classical approximation can be expressed entirely in terms
of the phase space weighting functions $U_\mu(\vec{k},\vec{x}) 
= N_\mu (\vec{x}) \, P(\vec{k})/[1+N(\vec{x}) \, P(\vec{k})]$,
where $N_\mu (\vec{x})=\bar{n}_\mu B_\mu^2$ is the effective density of
the tracer species $\mu$. These weighting functions are basically the FKP pair-weights.
These results make the case that, just as $P(\vec{k})$ is the density
of modes in Fourier space, and $N(\vec{x})=\sum_\mu N_\mu$ 
plays the role of the total effective density of tracers in position space, the quantity
$\frac12 U^2(\vec{k},\vec{x})$ can be interpreted as the density of information in phase 
space. The Fisher information matrix for the power spectrum is 
simply the sum (or trace) of 
the information over position-space volume [i.e., the effective volume $V_{eff}(\vec{k})$], 
and the Fisher matrix for the bias is the sum (or trace) of the information over 
the Fourier-space volume [what I have called here $\tilde{V}_{eff}(\vec{x})$].
The elements of the Fisher matrix which mix the power spectrum estimation
at $\vec{k}$ with the bias estimation at $\vec{x}$ are given simply in terms of $\frac12 U^2$
(the density of information), times the phase space volume occupied by the bins at
$\vec{k}$ and at $\vec{x}$. In a forthcoming paper (Abramo 2011, to appear) I will 
show how to use this result to jointly estimate the power spectrum and the bias (together
with the matter growth function) from the same dataset, without introducing hidden 
priors; and conversely, how to properly include priors in these estimations.


\vskip 0.3cm

\noindent {\it  Acknowledgements} -- 
I would like to thank Ravi Sheth for bringing to my attention some of the puzzles
that ultimately led to this work; many thanks also to G. Bernstein, Y.-C. Cai, A. Hamilton,
B. Jain and M. Strauss for useful comments and/or discussions.
I would like to thank the Department of Physics and Astronomy at the University of
Pennsylvannia, as well as the Department of
Astrophysical Sciences at Princeton University, for their warm hospitality.
This work was supported by both FAPESP and CNPq of Brazil.

\bibliographystyle{mn2e.bst}

\begin{thebibliography}{99}

\bibitem[\protect\citeauthoryear{Abbott et~al.}{2005}]{Abbott:2005bi}
Abbott, T. {\em et~al.}, 2005, astro-ph/0510346.

\bibitem[\protect\citeauthoryear{Abell et~al.}{2009}]{LSST:2009pq}
Abell, P. {\em et~al.}, 2009, 0912.0201

\bibitem[\protect\citeauthoryear{Abramo et~al.}{2011}]{2011arXiv1108.2657A}
Abramo, L.~R. {\em et~al.}, 2011, 1108.2657

\bibitem[\protect\citeauthoryear{Adelman-McCarthy et~al.}{2008a}]{adelman-mccarthy_sdss_2008}
{Adelman-McCarthy}, J.~K.  {\em et~al.}, 2008a, {VizieR} Online Data Catalog {\bf 2282}, 0

\bibitem[\protect\citeauthoryear{Adelman-McCarthy et~al.}{2008b}]{adelman-mccarthy_sixth_2008}
{Adelman-McCarthy}, J.~K.  {\em et~al.}, 2008b, \apjs {\bf 175}, 297

\bibitem[\protect\citeauthoryear{Albrecht et~al.}{2009}]{Albrecht09}
Albrecht, A. {\em et~al.}, 2009, 0901.0721

\bibitem[\protect\citeauthoryear{Ben\'{\i}tez et~al.}{2009}]{Benitez:2008fs}
{Ben{\'{\i}}tez}, N. {\em et~al.}, 2009, \apj {\bf 691}, 241


\bibitem[\protect\citeauthoryear{Bernstein}{1994}]{1994ApJ...424..569B}
Bernstein, G.~M. 1994, Astroph. J. {\bf 424}, 569

\bibitem[\protect\citeauthoryear{Blake \& Glazebrook}{2003}]{blake_probing_2003}
Blake, C., Glazebrook, K., 2003 \apj {\bf 594}, 665

\bibitem[\protect\citeauthoryear{Blake et~al.}{2011}]{2011MNRAS.415.2876B}
{Blake}, C. {\em et~al.}, 2011, \mnras {\bf 415}, 2876

\bibitem[\protect\citeauthoryear{BOSS}{}]{BOSS}
BOSS: http://cosmology.lbl.gov/boss/ .

\bibitem[\protect\citeauthoryear{Cai et~al.}{2011}]{2011MNRAS.412..995C}
{Cai}, Y.-C., {Bernstein}, G.~M., and {Sheth}, R.~K., 2011,
\mnras {\bf 412}, 995 (2011)

\bibitem[\protect\citeauthoryear{Cole et~al.}{2005}]{cole_2df_2005}
{Cole}, S. {\em et~al.}, 2005, \mnras {\bf 362}, 505

\bibitem[\protect\citeauthoryear{Eisenstein et~al.}{1999}]{eisenstein_cosmic_1998}
Eisenstein, D.~J., Hu, W., Tegmark, M., 1999 \apj {\bf 518}, 2

\bibitem[\protect\citeauthoryear{Feldman et~al.}{1994}]{1994ApJ...426...23F}
{Feldman}, H.~A., {Kaiser}, N., {Peacock}, J.~A., 1994, \apj {\bf 426}, 23


\bibitem[\protect\citeauthoryear{Hamilton}{1997a}]{Hamilton_97_I}
{Hamilton}, A.~J.~S., 1997a, \mnras {\bf 289}, 285

\bibitem[\protect\citeauthoryear{Hamilton}{1997b}]{Hamilton_97_II}
{Hamilton}, A.~J.~S., 1997b, \mnras {\bf 289}, 295

\bibitem[\protect\citeauthoryear{Hamilton}{1997c}]{hamilton_linear_1997}
Hamilton, A.~J.~S., 1997c, astro-ph/9708102, in {"The} Evolving Universe" 
ed. D. Hamilton, Kluwer Academic, p. 185-275 (1998).

\bibitem[\protect\citeauthoryear{Hamilton}{2005a}]{HamiltonRev05a}
Hamilton, A.~J.~S., 2005a, astro-ph/0503603, 
Lect. Notes Phys. {\bf 665}, 415.

\bibitem[\protect\citeauthoryear{Hamilton}{2005b}]{HamiltonRev05b}
Hamilton, A.~J.~S., 2005b, astro-ph/0503604, 
Lect. Notes Phys. {\bf 665}, 433.

\bibitem[\protect\citeauthoryear{Hamilton \& Culhane}{1996}]{1996MNRAS.278...73H}
{Hamilton}, A.~J.~S., {Culhane}, M., 1996, \mnras {\bf 278}, 73

\bibitem[\protect\citeauthoryear{McDonald \& Seljak}{2008}]{McDonald:2008sh}
McDonald, P., Seljak, U., 2009, JCAP {\bf 0910}, 007

\bibitem[\protect\citeauthoryear{Mukhanov}{2005}]{Mukhanov:05}
Mukhanov, V., {\em Physical Foundations of Cosmology} (Cambridge University Press,
 2005).

\bibitem[\protect\citeauthoryear{Norberg et~al.}{2002}]{2002MNRAS.332..827N}
{Norberg}, P., {\em et~al.}, 2002, \mnras {\bf 332}, 827

\bibitem[\protect\citeauthoryear{PAN-STARRS}{}]{PAN-STARRS}
PAN-STARRS: http://pan-starrs.ifa.hawaii.edu/public/ .

\bibitem[\protect\citeauthoryear{Percival et~al.}{2003}]{Percival:2003pi}
Percival, W.~J., Verde, L., Peacock, J.~A., 2004, \mnras {\bf 347}, 645

\bibitem[\protect\citeauthoryear{Percival et~al.}{2010}]{2010MNRAS.401.2148P}
{Percival}, W.~J., {\em et~al.}, 2010, \mnras {\bf 401}, 2148

\bibitem[\protect\citeauthoryear{Peres}{2002}]{PeresBook}
Peres, A. {\em Quantum Theory: Concepts and Methods} (Kluwer Academic
  Publishers, 2002).

\bibitem[\protect\citeauthoryear{Peter \& Uzan}{2009}]{Peter:2009zzc}
Peter, P., Uzan, J.-P., {\em Primordial Cosmology} (Oxford Univ. Press, 2009).

\bibitem[\protect\citeauthoryear{Sawangwit et~al.}{2011}]{2011arXiv1108.1198S}
{Sawangwit}, U., {\em et~al.}, 2011, 1108.1198.

\bibitem[\protect\citeauthoryear{Scoville et~al.}{2007}]{scoville_cosmic_2007}
Scoville, N., {\em et~al.}, 2007, \apjs {\bf 172}, 1.

\bibitem[\protect\citeauthoryear{Seljak et~al.}{2005a}]{Seljak:2004xh}
Seljak, U., {\em et~al.}, 2005a, Phys. Rev. {\bf D71}, 103515

\bibitem[\protect\citeauthoryear{Seljak et~al.}{2005b}]{2005PhRvD..71d3511S}
{Seljak}, U., {\em et~al.}, 2005b, Phys. Rev. {\bf D71}, 043511

\bibitem[\protect\citeauthoryear{Seo \& Eisenstein}{2003}]{seo_probing_2003}
Seo, H.-J., Eisenstein, D.~J., 2003 \apj {\bf 598}, 720

\bibitem[\protect\citeauthoryear{SUMIRE}{}]{SUMIRE}
SUMIRE: http://sumire.ipmu.jp/en/ .

\bibitem[\protect\citeauthoryear{Tegmark}{1997}]{Tegmark_Surveys_1997}
Tegmark, M., 1997, Phys. Rev. Lett. {\bf 79}, 3806.

\bibitem[\protect\citeauthoryear{Tegmark et~al.}{1997}]{Tegmark:1996bz}
Tegmark, M., Taylor, A., Heavens, A., 1997, \apj {\bf 480}, 22

\bibitem[\protect\citeauthoryear{Tegmark et~al.}{1998}]{1998ApJ...499..555T}
{Tegmark}, M., {Hamilton}, A.~J.~S., {Strauss}, M.~A., {Vogeley}, M.~S. ,
 {Szalay}, A.~S., 1998, \apj {\bf 499}, 555

\bibitem[\protect\citeauthoryear{Tegmark et~al.}{2004a}]{Tegmark:2003uf}
Tegmark, M., {\em et~al.}, 2004a, \apj {\bf 606}, 702

\bibitem[\protect\citeauthoryear{Tegmark et~al.}{2004b}]{tegmark_3d_2003}
{Tegmark}, M., {\em et~al.}, 2004b, \apj {\bf 606}, 702

\bibitem[\protect\citeauthoryear{Tegmark et~al.}{2004c}]{Tegmark:2003ud}
Tegmark, M., {\em et~al.}, 2004c, Phys. Rev. {\bf D69}, 103501


\bibitem[\protect\citeauthoryear{Tegmark et~al.}{2006}]{Tegmark:2006az}
Tegmark, M., {\em et~al.}, 2006, Phys. Rev. {\bf D74}, 123507

\bibitem[\protect\citeauthoryear{Vogeley \& Szalay}{1996}]{1996ApJ...465...34V}
{Vogeley}, M.~S.,  {Szalay},  A.~S., 1996, \apj {\bf 465}, 34

\bibitem[\protect\citeauthoryear{Wigner}{1932}]{Wigner32}
Wigner, E., 1932, Phys. Rev. {\bf 40}, 749.

\bibitem[\protect\citeauthoryear{White et~al.}{2008}]{White:2008jy}
White, M., Song, Y.-S.,  Percival, W.~J., 2008, \mnras {\bf 397}, 1348

\bibitem[\protect\citeauthoryear{York et~al.}{2000}]{York:2000gk}
York, D.~G., {\em et~al.}, 2000, \aj {\bf 120}, 1579

\end{thebibliography}

\end{document}